\documentclass[preprint,12pt]{elsarticle}

\usepackage{graphicx}

\usepackage{lineno}

\usepackage{amsmath}
\usepackage{bm}

\usepackage{float}

\usepackage{geometry}
\usepackage{threeparttable}
\biboptions{numbers,sort&compress}
\usepackage{subfigure}

\usepackage[colorlinks, citecolor=red]{hyperref}
\usepackage{xcolor}  

\journal{Elsevier}
\usepackage{CJK} 
\begin{document}

\begin{frontmatter}


\title{Explicit form of simplified Grad's 13 moments distribution function-based moment gas kinetic solver with unstructured meshes for the multiscale rarefied flow}



\author[rvt,rvtt]{W.~Liu}
\author[rvt]{C.~Shu\corref{cor1}}
\ead{mpeshuc@nus.edu.sg}
\author[rvttt]{Z.J.~Liu}
\cortext[cor1]{Corresponding author}

\address[rvt]{Department of Mechanical Engineering, National University of Singapore, 10 Kent Ridge Crescent, Singapore 119260}
\address[rvtt]{Division of Emerging Interdisciplinary Areas, The Hong Kong University of Science and Technology, Clear Water Bay, Hong Kong, China}
\address[rvttt]{Department of Mechanics and Aerospace Engineering, Southern University of Science and Technology, Shenzhen 518055, China}

\begin{abstract}

It is essential to efficiently solve multiscale flows covering the continuum regime to the rarefied regime. The explicit form of Grad’s 13 moments distribution function-based moment gas kinetic solver (G13-MGKS) has been proposed in our previous work [Comput. Math. Appl., 137 (2023), pp. 112–125], which demonstrates the potential for efficiently simulating continuum flows accurately and presenting reasonable predictions for rarefied flows at moderate Knudsen numbers on structured meshes. To further extend the solver's applicability to unstructured meshes, we propose the simplified version of the Grad's 13 moments distribution function-based moment gas kinetic solver (SG13-MGKS) with an explicit form of the numerical flux in the present paper. The Shakhov collision model has been adopted and validated within the framework of SG13-MGKS to ensure the correct Prandtl number in the simulation. Additionally, a simplified treatment for the numerical fluxes has been adopted to minimize the need for complex calculations of the gradient of integral coefficients. The performance of SG13-MGKS has been evaluated in numerical cases of Couette flow with temperature differences, flow passing through a NACA0012 airfoil, and pressure-driven flow in a variable-diameter circular pipe. Our results demonstrate that SG13-MGKS can achieve reasonably accurate computational results at Knudsen numbers below 0.2. Benefiting from the avoidance of discretization in velocity space, G13-MGKS is able to be two orders of magnitude faster compared to the conventional discrete velocity method. Furthermore, the simplified form of numerical fluxes and the fewer gradients of integration coefficients enable the performance of SG13-MGKS on unstructured grids with a saving of about 4 times the computation time and 3 times the memory cost compared to the previous version of G13-MGKS.

\end{abstract}

\begin{keyword}
Rarefied flow  \sep Boltzmann equation \sep Moment method \sep Finite volume method


\end{keyword}

\end{frontmatter}


\section{INTRODUCTION}
\label{S:1}

Rarefied gas flows are encountered across diverse industrial scenarios, including the aerospace vehicle \cite{li_gas-kinetic_2009,li_gas-kinetic_2011, chen_three-dimensional_2020}, photo lithography industry \cite{wang_investigation_2022,su_rarefaction_2017} and nuclear fusion devices \cite{zeng_general_2023-1, tantos_deterministic_2020}. Conventional numerical simulations based on the Navier-Stokes (NS) equations, grounded in the continuum hypothesis, face limitations in accurately capturing flow phenomena under enhanced rarefaction effects \cite{liu_simplified_2023, liu_numerical_2023}. Consequently, researchers are exploring the more fundamental Boltzmann equation to bridge this gap. In contrast to the conventional Computational Fluid Dynamics (CFD) solver relying on macroscopic variables, the Boltzmann equation adopts a statistical mechanics perspective, modeling gas behavior through the velocity distribution function of gas molecules, enabling a multiscale description from rarefied to continuum regimes \cite{wu_non-equilibrium_2016}.

Over the past few decades, significant progress has been made in numerical methods for solving the Boltzmann equation \cite{xu_unified_2021-1}. Representative methods fall into two main categories: stochastic methods and deterministic methods \cite{zhu_unified_2019}. Among many stochastic methods, the Direct Simulation Monte Carlo (DSMC) \cite{bird_molecular_1994} has achieved considerable success in studying high-speed rarefied flows. DSMC employs virtual gas molecule ensembles to simulate the collisions and transportation in molecular velocity space. Despite its nature of Lagrangian description, DSMC is challenged by the statistical noise significantly, particularly impacted in the low-speed flow simulations. In contrast, deterministic methods discretize the molecular velocity space by the discrete mesh, operating noise-free simulations based on the Eulerian perspective. As the most prominent deterministic approach, Discrete velocity method (DVM) \cite{goldstein_investigations_1989, harris_solution-adaptive_2021} performs well in low-speed flow simulations. However, the additional velocity space discretization substantially increases computational demands. Tests indicate that under the same physical mesh, DVM requires significantly more computations compared to the conventional CFD solver, hindering its practical applicability. 

To address rarefied effect beyond continuum assumptions and avoid extra discretization of the velocity space, Grad \cite{grad_kinetic_1949, grad_statistical_1952} proposed a set of truncated velocity distribution functions. These truncated distribution functions express the unknown distribution as the polynomial form of moment variables such as density, velocity, temperature, stress, and heat flux. Building upon these moments of the distribution function, Grad derived the moment method, showing promising efficiency in moderately rarefied flows compared to the conventional CFD method. Subsequently, Struchtrup et al \cite{struchtrup_macroscopic_2005} and Gu et al. \cite{gu_high-order_2009} proposed the regularized approach to involve the constitutive models into the the higher order moments. However, the governing equations in the moment method involve complex sets of partial differential equations and gradients of numerous macroscopic and higher-order quantities, potentially impacting computational complicity and stability \cite{wu_accuracy_2020}.

In our previous work, we amalgamated concepts from the approaches of DVM and moment method, specifically, we derive the governing equations of stress and heat flux from the discretized Boltzmann equation under finite-volume framework. The Grad's 13 moments distribution function is employed to approximate the unknown distribution function at the surrounding points of the cell interface. Subsequently, we performed explicit integration to obtain the expressions of numerical flux. This method is named as Grad’s 13 moments distribution function-based moment gas kinetic solver (G13-MGKS) \cite{liu_grads_2023}, and its effectiveness was verified in moderately rarefied flows on structured meshes, yielding reasonably accurate results. Moreover, previous testing indicated that G13-MGKS achieves computational efficiency approximately 3\% of that attained by DVM.

In the present study, we undertake a further step by extending the applicability of G13-MGKS to unstructured meshes. In comparison to prior versions, we have refined the calculation of interface fluxes by adopting a simpler upwind scheme. This refinement eliminates the need for calculating high-order integration coefficients and corresponding gradient at cell centers. Within this simplified version of G13-MGKS, the numerical fluxes of governing equations are deduced solely by computing gradients of macroscopic variables, stress, and heat flux. This simplification ensures that G13-MGKS remains efficient on unstructured meshes. Furthermore, in contrast to the previous version derived using the Bhatnagar-Gross-Krook (BGK) relaxation model \cite{chapman_mathematical_1962, pekardan_rarefaction_2018}, we introduce the Shakhov collision model \cite{shakhov_generalization_1968} to ensure that the present simplified form of G13-MGKS can accurately accommodate arbitrary Prandtl numbers. 

In the context of this paper, we conduct a series of numerical cases to verify the present simplified Grad’s 13 moments distribution function-based moment gas kinetic solver (SG13-MGKS), including the Couette flow with temperature difference, flow past a NACA0012 airfoil and pressure-driven pipe flow. The remainder of this paper is arranged as the following: Section \ref{SS:2-1} gives a quick review of the discretized Boltzmann equation and corresponding governing equation. Then, the explicit forms of the simplified numerical fluxes are given in section \ref{SS:2-2} and the procedure of the algorithm is summarized in section \ref{SS:2-3}. The numerical cases are studied in Section \ref{S:3}. Section \ref{S:4} summarizes this paper. 

\section{METHODOLOGY}
\label{S:2}

\subsection{\emph{Discretized Boltzmann equation and macroscopic governing equation}}
\label{SS:2-1}

Discrete velocity method (DVM) adopts the finite volume method (FVM) to discretize the Boltzmann-BGK equation \cite{xu_unified_2010, yang_improved_2018, liu_further_2024} within a discretized time step $\Delta t=t^{n+1}-t^{n}$ and the center of a discrete cell $i$,

\begin{equation}
f_{i}^{n+1}=f_{i}^{n}-\frac{\Delta t}{\left|\Omega_{i}\right|} \sum_{j \in N(i)}\left(\bm{\xi} \cdot \mathbf{n}_{i j} f_{i j}\left(\mathbf{x}_{i j}, \bm{\xi}, t\right)\right)\left|S_{i j}\right|+\frac{\Delta t}{2}\left(\frac{g_{+,i}^{n+1}-f_{i}^{n+1}}{\tau^{n+1}}+\frac{g_{+,i}^{n}-f_{i}^{n}}{\tau^{n}}\right),
\label{eq1}
\end{equation}

\noindent where $f(\mathbf{x}, \bm{\xi}, t)$ represent the gas velocity distribution function (VDF), which relate to the physical space $\mathbf{x}=(x, y, z)^{T}$, molecular velocity space $\bm{\xi}=\left(\xi_{x}, \xi_{y}, \xi_{z}\right)^{T}$ and the time $t$. The subscript $i j$ represents the relationship between cell $i$ and the neighboring cell $j$. $\left|S_{i j}\right|$ and $\mathbf{n}_{i j}$ represent the area and the unit normal vector of the cell interface. 

The trapezoidal rule are adopted for the approximation of the collision term \cite{xu_unified_2010, grunfeld_time_2014}. The ratio of the dynamic viscosity to the pressure determines the mean relaxation time, i.e., $\tau=\mu / p$. The modified equilibrium state $g_{+}(\mathbf{x}, \bm{\xi}, t)$ is given by,

\begin{equation}
g_{+}=g\left[1-(1-\operatorname{Pr}) \frac{\mathbf{q} \cdot \mathbf{C}}{p R_g T}\left(1-\frac{\mathbf{C}^2}{5 R_g T}\right)\right],
\label{eq2}
\end{equation}

\noindent where the  $\mathbf{C}=\bm{\xi}-\mathbf{U}$ denotes the peculiar velocity and Maxwellian equilibrium state $g(\mathbf{x}, \bm{\xi}, t)$ is 

\begin{equation}
g(\mathbf{x}, \bm{\xi}, t)=\frac{\rho}{\left(2 \pi R_{g} T\right)^{D / 2}} \exp \left[-\frac{\mathbf{C}^{2}}{2 R_{g} T}\right],
\label{eq3}
\end{equation}

\noindent Here, $R_{g}$ denotes the specific gas constant and $\operatorname{Pr}$ is the Prandtl number. The macroscopic quantities $\mathbf{W}=(\rho, \rho \mathbf{U}, \rho E)^{T}$, the stress tensor $\boldsymbol{\sigma}$ and heat flux $\mathbf{q}$ can be computed associated with the moment integral of the VDF as

\begin{equation}
\mathbf{W}=(\rho, \rho \mathbf{U}, \rho E)^{T}=\langle\psi f\rangle,
\label{eq4}
\end{equation}

\begin{equation}
\boldsymbol{\sigma}=\left\langle\left(\mathbf{C C}-\delta C^{2} / 3\right) f\right\rangle,
\label{eq5}
\end{equation}

\begin{equation}
\mathbf{q}=\left\langle\mathbf{C} C^{2} f/2\right\rangle,
\label{eq6}
\end{equation}

\noindent in which $\rho, \mathbf{U}=(U, V, W)$ and $T$ denote the density, the macroscopic velocity and the temperature, respectively. The symbol $\langle\cdot\rangle=\int_{-\infty}^{+\infty} \cdot \; d\bm{\xi}$ denotes the moment integral over the entire molecular velocity space and $\psi=\left(1, \bm{\xi}, \bm{\xi}^{2}/2 \right)^{T}$ represents the moment vector. 

Based on the conservation law in a relaxation process, the relaxation term should satisfy the compatibility condition as $\langle\psi(g_{+}-f) / \tau\rangle=0$. Further conducting the moment integral of Eq. (\ref{eq1}), the conservative form of macroscopic equations is given as:

\begin{equation}
\mathbf{W}_{i}^{n+1}=\mathbf{W}_{i}^{n}-\frac{\Delta t}{\left|\Omega_{i}\right|} \sum_{j \in N(i)} \mathbf{n}_{i j} \cdot \mathbf{F}_{i j}^{n+1}\left|S_{i j}\right|,
\label{eq7}
\end{equation}

\noindent where the corresponding numerical fluxes are given as

\begin{equation}
\begin{aligned}
\mathbf{F}_{i j}=\left\langle\bm{\xi} \psi f_{i j}\left(\mathbf{x}_{i j}, \bm{\xi}, t\right)\right\rangle.
\end{aligned}
\label{eq8}
\end{equation}


Following the treatment in the framework of previous version of G13-MGKS \cite{liu_grads_2023}, the governing equations of stress $\boldsymbol{\sigma}_{i}$  and heat flux $\mathbf{q}_{i}$ at the cell center can be obtained by conducting the moment integral related to the stress and heat flux on Eq. (\ref{eq1}) as

\begin{equation}
\boldsymbol{\sigma}_i^{n+1}=\left(1+\frac{\Delta t}{2\tau^{n+1}}\right)^{-1}\left[\boldsymbol{\sigma}_i^n-\frac{\Delta t}{\left|\Omega_i\right|} \sum_{j \in N(i)} \mathbf{n}_{i j} \cdot \mathbf{G}_{i j}^{n+1}\left|S_{i j}\right|\right],
\label{eq9}
\end{equation}

\noindent and 

\begin{equation}
\mathbf{q}_{i}^{n+1}=\left(1+\frac{\Delta t}{2\tau^{n+1}}\right)^{-1}\left[\mathbf{q}_{i}^{n}-\frac{\Delta t}{\left|\Omega_i\right|} \sum_{j \in N(i)} \mathbf{n}_{i j} \cdot\mathbf{H}_{i j}^{n+1}\left|S_{i j}\right|+\frac{\Delta t}{2}\left(\frac{{\mathbf{q}_{+, i}^{n+1}}}{\tau^{n+1}}+\frac{{\mathbf{q}_{+, i}^n}}{\tau^n}\right)\right],
\label{eq10}
\end{equation} 

\noindent where the numerical fluxes related to the stress $\mathbf{G}_{i j}$ and heat flux $\mathbf{H}_{i j}$ could be defined as

\begin{equation}
\begin{aligned}
\mathbf{G}_{i j}=\left\langle\bm{\xi}\left(\mathbf{C}_{i}\mathbf{C}_{i}-\delta \mathbf{C}_{i}^{2} / 3\right) f_{i j}\right\rangle.
\end{aligned}
\label{eq11}
\end{equation}

\noindent and

\begin{equation}
\begin{aligned}
\mathbf{H}_{i j}=\left\langle\bm{\xi}\mathbf{C}_{i}\mathbf{C}_{i}^{2} f_{i j}/2\right\rangle. 
\end{aligned}
\label{eq12}
\end{equation}

\noindent To associate the numerical flux related to stress and heat flux with the macroscopic variables, the peculiar velocities at the cell center $\mathbf{C}_{i}\left(\mathbf{x}_{i}\right)=\bm{\xi}-\mathbf{U}_{i}$ are utilized where $\mathbf{U}_{i}$ is the macroscopic velocities at the cell center.

The additional source term for the heat flux $\mathbf{q}_{+, i}$ is defined as

\begin{equation}
\begin{aligned}
\mathbf{q}_{+, i}=\left\langle{\mathbf{C}}_{i} {\mathbf{C}}_{i}^2 g_{+, i} / 2\right\rangle=(1-\operatorname{Pr}) \mathbf{q}_{i}.
\end{aligned}
\label{eq13}
\end{equation}
Then the heat flux can be updated at the cell center as

\begin{equation}
\mathbf{q}_i^{n+1}=\left(1+\operatorname{Pr} \frac{\Delta t}{2 \tau^{n+1}}\right)^{-1}\left[\left(1+(1-\operatorname{Pr}) \frac{\Delta t}{2 \tau^n}\right) \mathbf{q}_i^n-\frac{\Delta t}{\left|\Omega_i\right|} \sum_{j \in N(i)} \mathbf{n}_{i j} \cdot \mathbf{H}_{i j}^{n+1}\left|S_{i j}\right|\right].
\label{qi_new}
\end{equation} 

It is worth noting that the updating of the above governing equations for macroscopic quantities, stress and heat flux all depend on the accurate calculation of the macroscopic numerical fluxes at the interface. In conventional DVM, the numerical flux at the interface can be obtained using numerical integration based on the distribution function $f_{i j}$. Therefore, the velocity distribution function (VDFs) need to be discretized in the velocity space and evolved according to the Eq. (\ref{eq1}) at the cell center. Then the VDFs can be interpolated from the cell center to the cell interface to obtian $f_{i j}$. However, the discretization of the velocity space as well as the evolution and numerical summation of the VDFs will consume huge computational time and memory.

\subsection{\emph {Explicit form of the simplified explicit numerical fluxes on the unstructured mesh}}
\label{SS:2-2}

To release the amount of computation and memory consumed by the evolution of VDFs, G13-MGKS adopts the local solution of Boltzmann-BGK equation at the cell interface as

\begin{equation}
\begin{aligned}
f_{i j}\left(\mathbf{x}_{i j}, \bm{\xi}, t\right)= &\frac{\Delta t}{\tau^{n+1}+\Delta t} g\left(\mathbf{x}_{i j}, \bm{\xi}, t^{n+1}\right)+\frac{\tau^{n+1}}{\tau^{n+1}+\Delta t} f\left(\mathbf{x}_{i j}-\bm{\xi} \Delta t, \bm{\xi}, t^n\right).
\end{aligned}
\end{equation}
In order to approximate the unknown initial VDFs at surrounding points around the interface $f\left(\mathbf{x}_{i j}-\bm{\xi} \Delta t, \bm{\xi}, t^n\right)$, the truncated distribution function proposed by Grad \cite{grad_kinetic_1949, grad_statistical_1952} has been utilized, which expresses the unknown initial VDF as an explicit function of macroscopic variables, stress and heat flux. Here, the Grad's distribution function for 13 moments (G13) \cite{grad_statistical_1952} could be given as

\begin{equation}
f^{G 13}=g\left(1+\frac{\boldsymbol{\sigma}}{2 p R_{g} T} \cdot \mathbf{C C}-\frac{\mathbf{q} \cdot \mathbf{C}}{p R_{g} T}\left(1-\frac{\mathbf{C}^{2}}{5 R_{g} T}\right)\right).
\label{eq14}
\end{equation}
This approximation helps the distribution function at the interface to obtain a complete explicit expression, but the fact that the initial distribution function is located at points around the interface will result in the computation of the numerical fluxes containing a large number of gradient calculations, which greatly reduces the efficiency of the performance of G13-MGKS on unstructured meshes. 

In order to avoid the increase in computational cost brought by the large number of gradients, we try to approximate the VDFs at the cell interface directly by using a truncated distribution function as

\begin{equation}
\begin{aligned}
f_{i j}\left(\mathbf{x}_{i j}, \boldsymbol{\xi}, t\right)=\operatorname{Hea}\left(\boldsymbol{\xi} \cdot \mathbf{n}_{i j}\right) f_{i j}^{G13}\left(\mathbf{x}_{i j}^{L}\right)+\left[1-\operatorname{Hea}\left(\boldsymbol{\xi}_k \cdot \mathbf{n}_{i j}\right)\right] f_{i j}^{G13}\left(\mathbf{x}_{i j}^{R}\right),
\end{aligned}
\end{equation}
\noindent where Hea denotes the Heaviside function determining which side of the cell center the VDF derives from.  The superscripts $L$ and $R$ represent variables at the left and right sides of the interface, respectively. It is readily  to notice that the reconstruction of numerical fluxes at the interface need only involve the macroscopic quantities, stress tensor and heat flux $\phi=(\mathbf{W}, \boldsymbol{\sigma}, \mathbf{q})^{T}$, which are needed to be interpolated to the cell interface as

\begin{equation}
\begin{aligned}
\phi\left(\mathbf{x}_{i j}^L\right)=\phi\left(\mathbf{x}_i\right)+V\left(\phi_i\right) \nabla \phi\left(\mathbf{x}_i\right) \cdot\left(\mathbf{x}_{i j}-\mathbf{x}_i\right), & \mathbf{n}_{i j} \cdot \bm{\xi} \geq 0, \\
\phi\left(\mathbf{x}_{i j}^R\right)=\phi\left(\mathbf{x}_j\right)+V\left(\phi_j\right) \nabla \phi\left(\mathbf{x}_j\right) \cdot\left(\mathbf{x}_{i j}-\mathbf{x}_j\right), & \mathbf{n}_{i j} \cdot \bm{\xi}<0,
\end{aligned}
\label{ls}
\end{equation}

\noindent where the $V\left(\phi\right)$ is the Venkatakrishnan’s limiter \cite{venkatakrishnan_convergence_1995} functions of the variable $\phi$. The gradient $\nabla \phi(\mathbf{x})^{n}$ can be computed by the the least squares method directly \cite{bjorck_least_1990, lm_yang_improved_2019}. 

In order to apply the reconstruction of numerical fluxes on the unstructured mesh, we need to transform variables such as macroscopic quantities, stresses and heat fluxes from the global to the local coordinate system, i.e., from the $x-y-z$ to the ${n}-{\tau}-s$ plane, where ${n}$, ${\tau}$ and $s$ are the normal and two tangential directions, respectively. Take the two-dimensional case as an example, suppose that its unit normal vector for a certain cell interface is $\mathbf{n}_{i j}=\left\{n_x, n_y\right\}$, the calculation of the relevant coordinate transformations are given as follows:

\begin{equation}
\begin{aligned}
& \xi_n=n_x \xi_x+n_y \xi_y, \\
& \xi_\tau=-n_y \xi_x+n_x \xi_y,
\end{aligned}
\label{eq15.1}
\end{equation}

\begin{equation}
\begin{aligned}
& U_n=n_x U+n_y V, \\
& U_\tau=-n_y U+n_x V,
\end{aligned}
\label{eq15.2}
\end{equation}

\begin{equation}
\begin{aligned}
& \sigma_{n n}=n_x^2 \sigma_{x x}+2 n_x n_y \sigma_{x y}+n_y^2 \sigma_{y y}, \\
& \sigma_{n \tau}=n_x n_y\left(\sigma_{y y}-\sigma_{x x}\right)+\left(n_x^2-n_y^2\right) \sigma_{x y},\\
& \sigma_{\tau \tau}=n_x^2 \sigma_{y y}-2 n_x n_y \sigma_{x y}+n_y^2 \sigma_{x x},
\end{aligned}
\label{eq15.3}
\end{equation}

\noindent and

\begin{equation}
\begin{aligned}
\begin{aligned}
& q_n=n_x q_x+n_y q_y, \\
& q_\tau=-n_y q_x+n_x q_y,
\end{aligned}
\end{aligned}
\label{eq15.4}
\end{equation}

\noindent where $\xi_n$ and $\xi_\tau$ denote the normal and tangential molecular velocities. $U_n$ and $U_\tau$ represents the normal and tangential macroscopic velocities. $\sigma$ and $q$ are the independent components in the stress tensor and heat flux. 

Here we define the integral in the velocity space with the interval from negative infinity to zero and the interval from zero to infinity as

\begin{equation}
\begin{aligned}
& \langle\cdots\rangle_{>0}=\frac{1}{\rho^L} \int_0^{+\infty} \int_{-\infty}^{+\infty} \int_{-\infty}^{+\infty}(\cdots) f_{i j}^{G 13}\left(\mathbf{x}_{i j}^L\right) d \zeta d \xi_\tau d \xi_n . \\
& \langle\cdots\rangle_{<0}=\frac{1}{\rho^R} \int_{-\infty}^0 \int_{-\infty}^{+\infty} \int_{-\infty}^{+\infty}(\cdots) f_{i j}^{G 13}\left(\mathbf{x}_{i j}^R\right) d \zeta d \xi_\tau d \xi_n,
\end{aligned}
\end{equation}
in which $\zeta$ is the phase energy to replace the $\xi_{s}$. Then the numerical fluxes based on the local coordinate system are calculated as follows, while the superscript * indicates that the corresponding numerical fluxes are in the local coordinate system. The explicit formulations of the numerical fluxes related to the initial state $\mathbf{F}_{i j}^{*}$ could be given as 

\begin{equation}
\begin{aligned}
\mathbf{F}_{i j}^*=\rho^L \mathbf{X}^L+\rho^R \mathbf{X}^R,
\end{aligned}
\label{eq16}
\end{equation}

\noindent where $\mathbf{X}^{L}$ and $\mathbf{X}^{R}$ are the integration parameters, which could be computed as

\begin{equation}
\mathbf{X}^L(1)=\left\langle\xi_n^1 \xi_\tau^0 \zeta^0\right\rangle_{>0}, \quad \mathbf{X}^R(1)=\left\langle\xi_n^1 \xi_\tau^0 \zeta^0\right\rangle_{<0}
\label{eq17}
\end{equation}

\begin{equation}
\mathbf{X}^L(2)=\left\langle\xi_n^2 \xi_\tau^0 \zeta^0\right\rangle_{>0}, \quad \mathbf{X}^R(2)=\left\langle\xi_n^2 \xi_\tau^0 \zeta^0\right\rangle_{<0}
\label{eq18}
\end{equation}

\begin{equation}
\mathbf{X}^L(3)=\left\langle\xi_n^1 \xi_\tau^1 \zeta^0\right\rangle_{>0}, \quad \mathbf{X}^R(3)=\left\langle\xi_n^1 \xi_\tau^1 \zeta^0\right\rangle_{<0}
\label{eq19}
\end{equation}

\noindent and

\begin{equation}
\begin{aligned}
& \mathbf{X}^L(4)=\frac{1}{2}\left(\left\langle\xi_n^3 \xi_\tau^0 \zeta^0\right\rangle_{>0}+\left\langle\xi_n^1 \xi_\tau^2 \zeta^0\right\rangle_{>0}+\left\langle\xi_n^1 \xi_\tau^0 \zeta^2\right\rangle_{>0}\right), \\
& \mathbf{X}^R(4)=\frac{1}{2}\left(\left\langle\xi_n^3 \xi_\tau^0 \zeta^0\right\rangle_{<0}+\left\langle\xi_n^1 \xi_\tau^2 \zeta^0\right\rangle_{<0}+\left\langle\xi_n^1 \xi_\tau^0 \zeta^2\right\rangle_{<0}\right),
\end{aligned}
\label{eq20}
\end{equation}

\noindent The calculation of moment integral $\left\langle\xi_{n}^{o} \xi_{\tau}^{p} \zeta^{q} \right\rangle_{>0}$ and $\left\langle\xi_{n}^{o} \xi_{\tau}^{p} \zeta^{q} \right\rangle_{<0}$ could be found in \ref{A:1}. 



Then the explicit formulations of the numerical fluxes related to stress $\mathbf{G}_{i j}^{*}$ and heat flux $\mathbf{H}_{i j}^{*}$ would be given in the remaining part of this section. To update the independent components in the stress tensor $\sigma_{n n}$, $\sigma_{n \tau}$ and $\sigma_{\tau \tau}$, the corresponding components of $G_{n n}, G_{n \tau}$ and $G_{\tau \tau}$ in the numerical flux $\mathbf{G}_{i j}^{*}$ could be expressed as  

\begin{equation}
\begin{aligned}
G_{n n}=\frac{1}{3}\left(2 m_{11}-m_{22}-m_{33}\right),
\end{aligned}
\label{eq21}
\end{equation}

\begin{equation}
\begin{aligned}
G_{n \tau}=m_{12},
\end{aligned}
\label{eq22}
\end{equation}

\begin{equation}
\begin{aligned}
G_{\tau \tau}=\frac{1}{3}\left(2 m_{22}-m_{11}-m_{33}\right),
\end{aligned}
\label{eq23}
\end{equation}

\noindent where the parameters related to the stress $m_{11}$, $m_{12}$, $m_{22}$ and $m_{33}$ could be given as

\begin{equation}
\begin{aligned}
m_{11} =&\rho^L \mathbf{Y}^{L}(1)+\rho^R \mathbf{Y}^{R}(1)-2 {U}_{i, n} \mathbf{F}_{i j}^{*}(2)+\left({U}_{i, n}\right)^{2}\mathbf{F}_{i j}^{*}(1),
\end{aligned},
\label{eq24}
\end{equation}

\begin{equation}
\begin{aligned}
m_{12} =&\rho^L \mathbf{Y}^{L}(2)+\rho^R \mathbf{Y}^{R}(2)-{U}_{i, \tau} \mathbf{F}_{i j}^{*}(2),
\end{aligned},
\label{eq25}
\end{equation}

\begin{equation}
\begin{aligned}
m_{22} =&\rho^L \mathbf{Y}^{L}(3)+\rho^R \mathbf{Y}^{R}(3)-2 {U}_{i, \tau} \mathbf{F}_{i j}^{*}(3)+\left({U}_{i, \tau}\right)^{2} \mathbf{F}_{i j}^{*}(1),
\end{aligned}
\label{eq26}
\end{equation}

\begin{equation}
\begin{aligned}
m_{33}=\rho^L \mathbf{Y}^{L}(4)+\rho^R \mathbf{Y}^{R}(4).
\end{aligned}
\label{eq27}
\end{equation}
Here, ${U}_{i, n}$ and ${U}_{i, \tau}$ are the macroscopic velocities at the cell center along the normal and tangential direction of the cell interface, which should be calculated based on the macroscopic variables $\mathbf{W}_{i}$ at the cell center. 

To update the components of heat flux $q_{n}$ and $q_{\tau}$, the corresponding numerical flux related to the heat flux $\mathbf{H}_{i j}^{*}=\left(H_{n}, H_{\tau}\right)^{T}$ could be expressed as  

\begin{equation}
\begin{aligned}
H_{n}=\frac{1}{2}\left(m_{111}+m_{122}+m_{133}\right),
\end{aligned}
\label{eq28}
\end{equation}

\begin{equation}
\begin{aligned}
H_{\tau}=\frac{1}{2}\left(m_{211}+m_{222}+m_{233}\right),
\end{aligned}
\label{eq29}
\end{equation}

\noindent where the parameters related to the heat flux $m_{111}$, $m_{122}$, $m_{133}$, $m_{211}$, $m_{222}$ and $m_{233}$ could be given as

\begin{equation}
\begin{aligned}
m_{111} =&\rho^L \mathbf{Z}^{L}(1)+\rho^R \mathbf{Z}^{R}(1)-3 {U}_{i, n}\left(m_{11}+2 {U}_{i, n} \mathbf{F}_{i j}^{*}(2)-\left({U}_{i, n}\right)^{2} \mathbf{F}_{i j}^{*}(1)\right)+3 {U}_{i, n} \mathbf{F}_{i j}^{*}(2) \\
&-\left({U}_{i, n}\right)^{3} \mathbf{F}_{i j}^{*}(1),
\end{aligned}
\label{eq30}
\end{equation}

\begin{equation}
\begin{aligned}
m_{122} =&\rho^L \mathbf{Z}^{L}(2)+\rho^R \mathbf{Z}^{R}(2)-2 {U}_{i, \tau}\left(m_{12}+{U}_{i, \tau} \mathbf{F}_{i j}^{*}(2)\right)-{U}_{i, n}\left(m_{22}+2 {U}_{i, \tau} \mathbf{F}_{i j}^{*}(3)-\left({U}_{i, \tau}\right)^{2} \mathbf{F}_{i j}^{*}(1)\right) \\
&+\left({U}_{i, \tau}\right)^{2} \mathbf{F}_{i j}^{*}(2)-2 {U}_{i, n} {U}_{i, \tau} \mathbf{F}_{i j}^{*}(3)+{U}_{i, n}\left({U}_{i, \tau}\right)^{2} \mathbf{F}_{i j}^{*}(1),
\end{aligned}
\label{eq31}
\end{equation}

\begin{equation}
\begin{aligned}
m_{133} =&\rho^L \mathbf{Z}^{L}(3)+\rho^R \mathbf{Z}^{R}(3)- {U}_{i, n} m_{33},
\end{aligned}
\label{eq32}
\end{equation}

\begin{equation}
\begin{aligned}
m_{211} =&\rho^L \mathbf{Z}^{L}(4)+\rho^R \mathbf{Z}^{R}(4)+\left({U}_{i, n}\right)^{2} \mathbf{F}_{i j}^{*}(3)+2 {U}_{i, n} {U}_{i, \tau} \mathbf{F}_{i j}^{*}(2)-\left({U}_{i, n}\right)^{2} {U}_{i, \tau} \mathbf{F}_{i j}^{*}(1),
\end{aligned}
\label{eq33}
\end{equation}

\begin{equation}
\begin{aligned}
m_{222} =&\rho^L \mathbf{Z}^{L}(5)+\rho^R \mathbf{Z}^{R}(5)-3 {U}_{i, \tau}\left(m_{22}+2 {U}_{i, \tau} \mathbf{F}_{i j}^{*}(3)-\left({U}_{i, \tau}\right)^{2} \mathbf{F}_{i j}^{*}(1)\right)+3\left({U}_{i, \tau}\right)^{2} \mathbf{F}_{i j}^{*}(3) \\
&-\left({U}_{i, \tau}\right)^{3} \mathbf{F}_{i j}^{*}(1),
\end{aligned}
\label{eq34}
\end{equation}

\begin{equation}
\begin{aligned}
m_{223} =&\rho^L \mathbf{Z}^{L}(6)+\rho^R \mathbf{Z}^{R}(6)-{U}_{i, \tau} m_{33}.
\end{aligned}
\label{eq35}
\end{equation}

The formulations of parameters including $\mathbf{Y}(1) \sim \mathbf{Y}(4)$ and  $\mathbf{Z}(1) \sim \mathbf{Z}(6)$ can be found in \ref{C:1}. Finally, the numerical fluxes of macroscopic equations in the local coordinate system are transformed to the numerical fluxes in the global coordinate system $\mathbf{F}_{i j}$ as

\begin{equation}
\begin{aligned}
\begin{aligned}
& \mathbf{F}_{i j}(1)=\mathbf{F}_{i j}^*(1), \\
& \mathbf{F}_{i j}(2)=n_x \mathbf{F}_{i j}^*(2)-n_y \mathbf{F}_{i j}^*(3), \\
& \mathbf{F}_{i j}(3)=n_y \mathbf{F}_{i j}^*(2)+n_x \mathbf{F}_{i j}^*(3), \\
& \mathbf{F}_{i j}(4)=\mathbf{F}_{i j}^*(4).
\end{aligned}
\end{aligned}
\label{eq35.1}
\end{equation}
The numerical fluxes for the independent component of $\mathbf{G}_{i j}=\left(G_{xx}, G_{xy}, G_{yy}\right)^{T}$ and $\mathbf{H}_{i j}=\left(H_{x}, H_{y}\right)^{T}$ in the global coordinate system are given as

\begin{equation}
\begin{aligned}
& G_{x x}=n_x^2 G_{n n}-2 n_x n_y G_{n \tau}+n_y^2 G_{\tau \tau}, \\
& G_{x y}=n_x n_y\left(G_{\tau \tau}-G_{n n}\right)+\left(n_x^2-n_y^2\right) G_{n \tau}, \\
& G_{y y}=n_x^2 G_{\tau \tau}+2 n_x n_y G_{n \tau}+n_y^2 G_{n n},
\end{aligned}
\label{eq35.2}
\end{equation}

\noindent and

\begin{equation}
\begin{aligned}
& H_x=n_x H_n-n_y H_\tau, \\
& H_y=n_y H_n+n_x H_\tau.
\end{aligned}
\label{eq35.3}
\end{equation}

\subsection{\emph{Computational sequence}}
\label{SS:2-3}

\begin{itemize}
\setlength{\itemsep}{0pt}
\setlength{\parsep}{0pt}
\setlength{\parskip}{0pt}
\item[1)]
Determine the time step based on the CFL condition \cite{cercignani_mathematical_1969}. 
\item[2)] 
Calculate derivatives of macroscopic variables, stress and heat flux by the  least squares method. Then interpolate these quantities to the cell interface by Eq. (\ref{ls}).
\item[3)]
Transforming the macroscopic quantities, stress and heat flux from the global coordinate to the local coordinate according to Eqs. (\ref{eq15.1}) - (\ref{eq15.4}).
\item[4)]
Calculate the numerical fluxes related to macroscopic variables $\mathbf{F}_{i j}^*$, stress $\mathbf{G}_{i j}^*$ and heat flux $\mathbf{H}_{i j}^*$ by Eq. (\ref{eq16}), Eqs. (\ref{eq21}) - (\ref{eq23}) and Eqs. (\ref{eq28}) - (\ref{eq29}), respectively.
\item[5)]
Transforming the numerical fluxes related to macroscopic variables, stress and heat flux from the local coordinate to the global coordinate according to Eqs. (\ref{eq35.1}) - (\ref{eq35.3}).
\item[6)]
Update the macroscopic governing equations for the macroscopic variables $\mathbf{W}_{i}^{n+1}$, stress $\boldsymbol{\sigma}_{i}^{n+1}$ and heat flux $\mathbf{q}_{i}^{n+1}$ by Eq. (\ref{eq7}), Eq. (\ref{eq9}) and Eq. (\ref{qi_new}), respectively.
\item[7)]
Repeat steps (1) – (6) before the convergence criterion can be satisfied.
\end{itemize}

\section{NUMERICAL EXPERIMENTS}
\label{S:3}

In this section, a number of numerical cases are utilized to verify the proposed method. The CFL number is uniformly set to $0.95$ for all the cases in the present section. In order to verify the accuracy of the results, we selected analytical solutions, direct simulation monte carlo (DSMC) method and improved discrete velocity method (IDVM) as reference results.

\subsection{\emph{Couette flow with temperature difference}}
\label{sec3-1}

\begin{figure}[H]
\centering
\includegraphics[width=10cm]{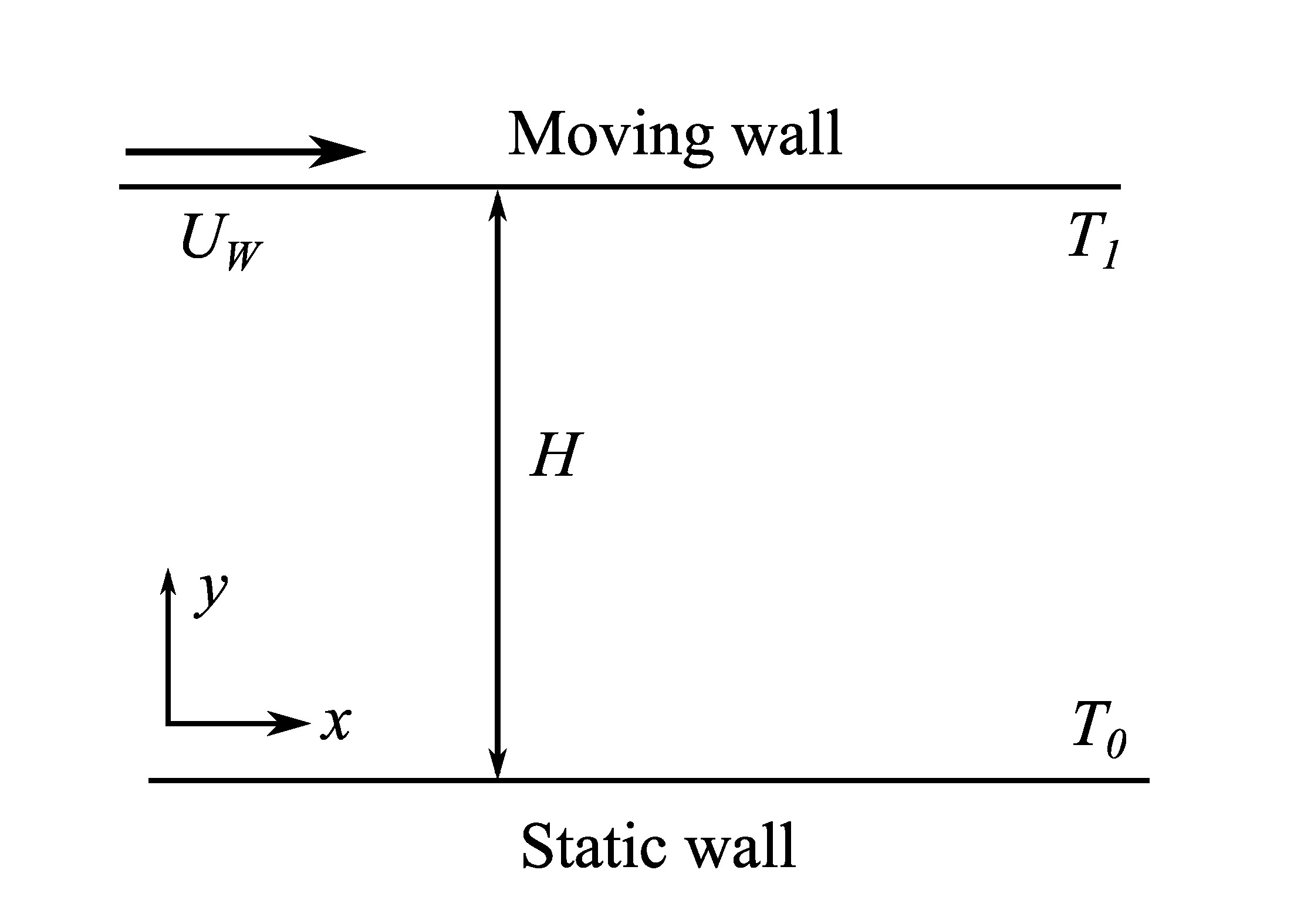}
\caption{Schematic of Couette flow with temperature difference.}
\label{Fig1}
\end{figure}

Couette flow is a classic benchmark case for verifying that an method can recover the correct Prandtl number. The test cases are designed in the following form: Two infinite flat plates are placed parallel to each other and separated by distance of $H = 1.0$ as depicted in Fig. \ref{Fig1}. The bottom wall is fixed with the temperature $T_{0}$, and the top plate is moving at a speed $U_{W}$ and the temperature of the top wall is $T_{1}$. The computational domain of $\left[0, 0.2\right]$ $\times$ $\left[0, 1\right]$ is uniformly discretized by $20$ $\times$ $100$ cells along the $x$ and $y$ directions, respectively. The left and right boundaries of the computational region are set to periodic conditions. The upper and lower flat plates are set as isothermal wall with the conditions of fully diffuse reflection. 

When the flow reaches a steady state, the temperature distribution can be obtained under the assumption of constant viscosity and heat conduction coefficients, which can be written as 

\begin{equation}
\begin{aligned}
& T_0=T_1, \quad T-T_0=\operatorname{Pr} \frac{U_{W}^2}{2 c_p} \frac{y}{H}\left(1-\frac{y}{H}\right) \\
& T_0 \neq T_1, \quad \frac{T-T_0}{T_1-T_0}=\frac{y}{H}+\frac{\operatorname{Pr} \cdot \mathrm{Ec}}{2} \frac{y}{H}\left(1-\frac{y}{H}\right)
\end{aligned},
\label{eq36}
\end{equation}

\noindent where $c_p = c_v + R$ is the specific heat capacity at constant pressure. $c_v$ and $R$ are the specific heat capacity at constant volume and the universal gas constant, respectively. The specific heat ratio $\gamma$ for the argon gas is 

\begin{equation}
\begin{aligned}
\gamma=\frac{c_v}{c_p}=\frac{K+4}{K+2}=5/3,
\end{aligned}
\label{eq37}
\end{equation}

\noindent where $K = 1$ is the internal degree of freedom of molecules for the two dimension. Eckert number is defined by

\begin{equation}
\begin{aligned}
\mathrm{Ec}=U_{W}^2 / c_p\left(T_1-T_0\right),
\end{aligned}
\label{eq38}
\end{equation}
The flow with reference temperature $T_{0}=1.0$, reference density $\rho = 1.0$ and reference velocity of $U_{W} = 1.0$ are aoopted in the present case. The dynamic viscosity $\mu$ should be calculated based on the Reynolds number as $\operatorname{Re} = 500$ for the continuum flow and the dynamic viscosity is given as $\mu=\rho U_{W} H / \operatorname{Re}$.

\begin{figure}[H]
\centering
\subfigure[]{
\includegraphics[width=4.6cm]{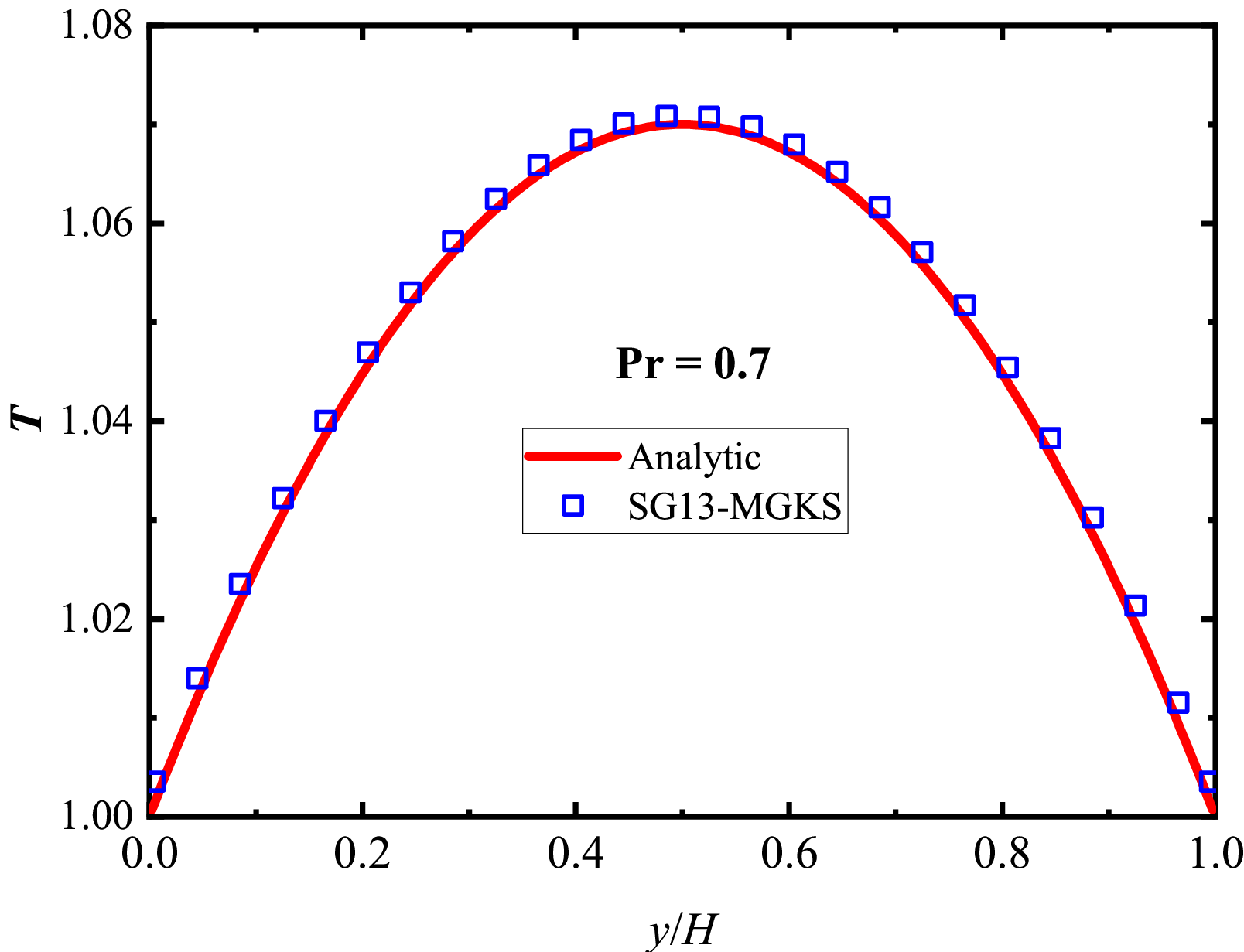}
}
\quad
\subfigure[]{
\includegraphics[width=4.6cm]{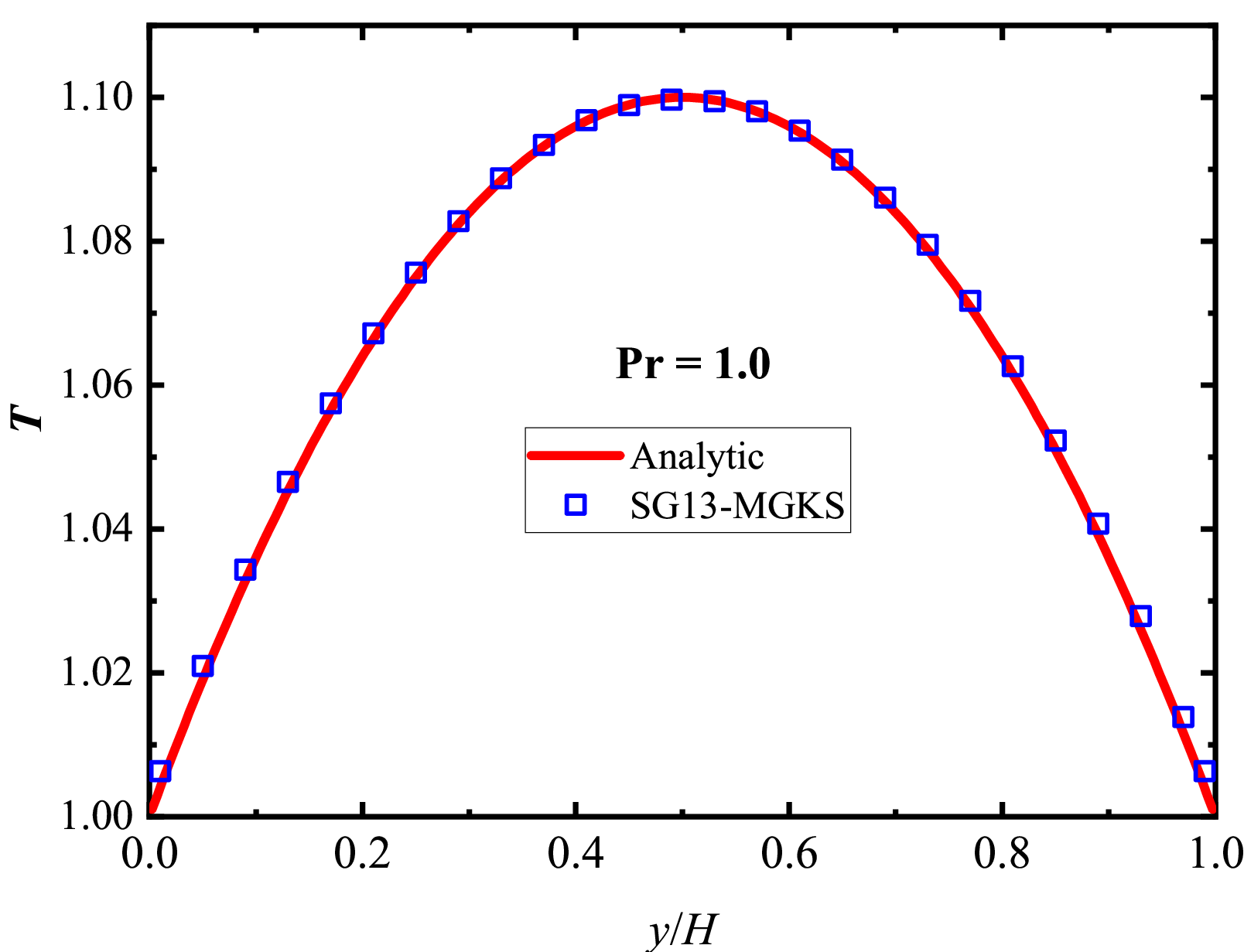}
}
\quad
\subfigure[]{
\includegraphics[width=4.6cm]{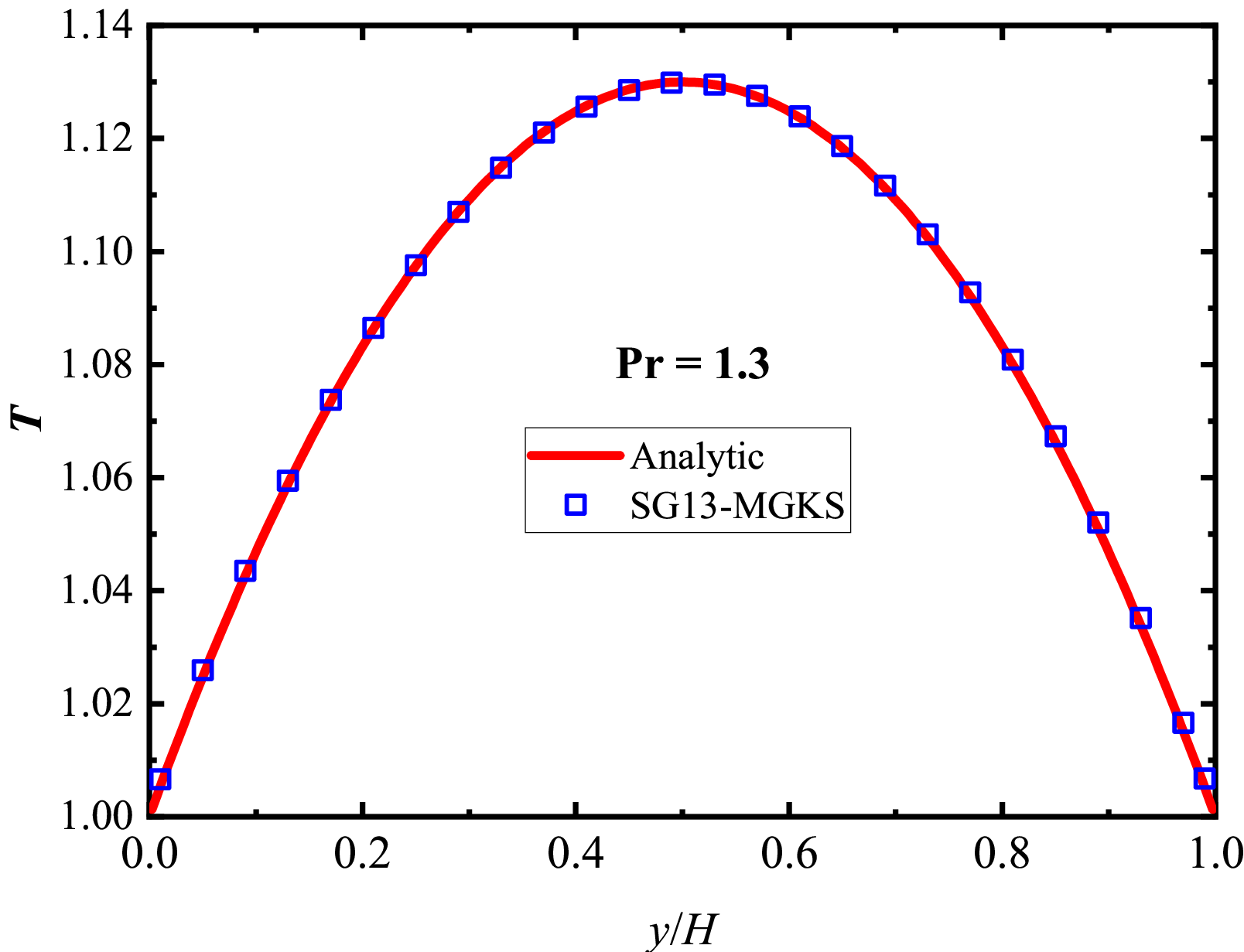}
}
\caption{Comparison of temperature profile for various Prandtl numbers when $T_{1}=T_{0}$, (a) $\operatorname{Pr}=0.7$, (b) $\operatorname{Pr}=1.0$ and (c) $\operatorname{Pr}=1.3$.}
\label{Fig2}
\end{figure}

\begin{figure}[H]
\centering
\subfigure[]{
\includegraphics[width=4.6cm]{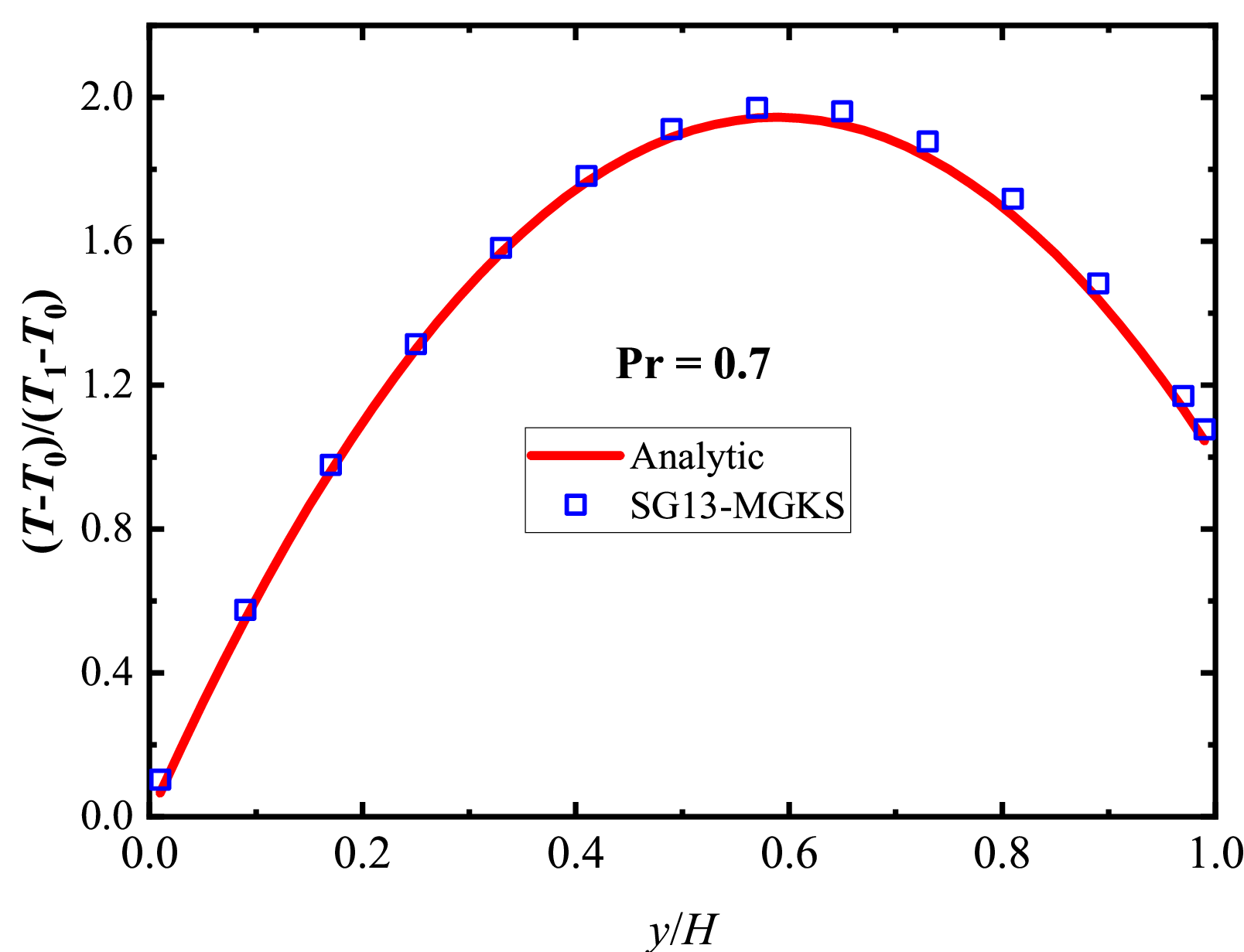}
}
\quad
\subfigure[]{
\includegraphics[width=4.6cm]{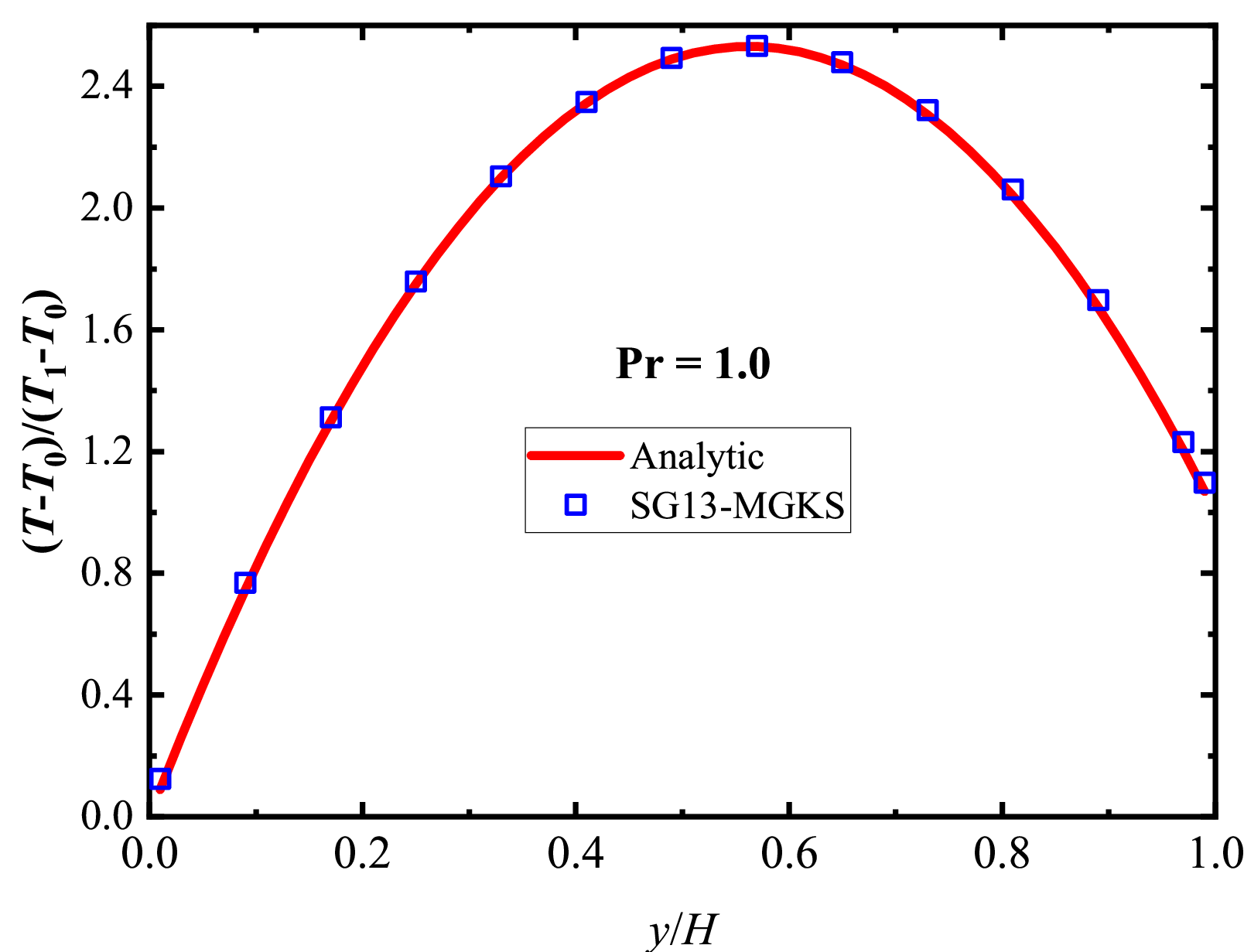}
}
\quad
\subfigure[]{
\includegraphics[width=4.6cm]{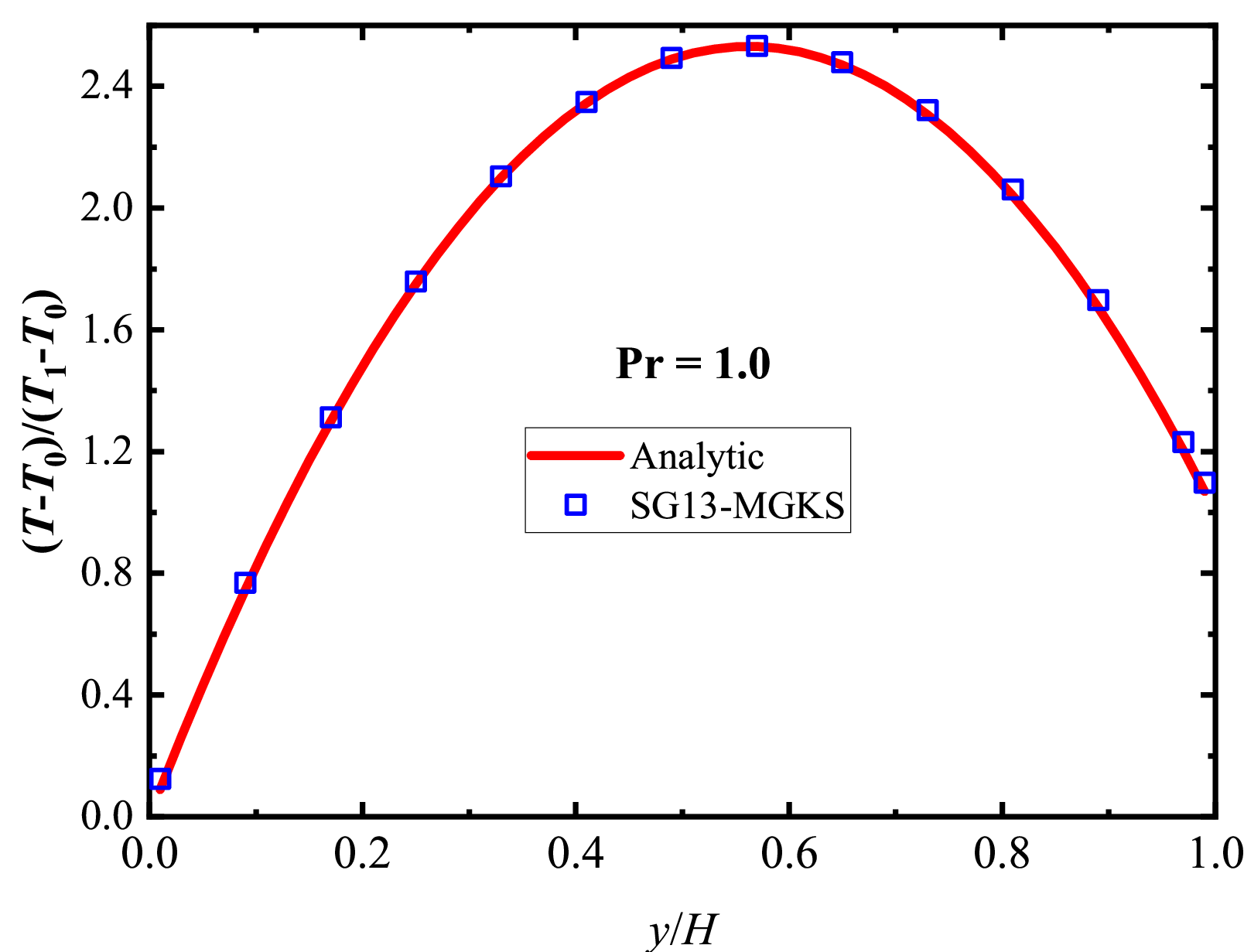}
}
\caption{Comparison of temperature profile for various Prandtl numbers when $T_{1}=1.05 T_{0}$, (a) $\operatorname{Pr}=0.7$, (b) $\operatorname{Pr}=1.0$ and (c) $\operatorname{Pr}=1.3$.}
\label{Fig3}
\end{figure}

Fig. \ref{Fig2} presents the temperature profiles of flow at different Prandtl numbers with $T_0=T_1$. In this case, the temperature profile would obey a parabolic distribution based on Eq. (\ref{eq36}). As shown in figures, the results obtained from the SG13-MGKS can match well with the analytic solutions at $\operatorname{Pr}=0.7, 1.0$ and $1.3$. When the temperature is increased to $T_1= 1.05 T_0$, the results in Fig. \ref{Fig3} give the distribution of the temperature. When the temperature of the two plates differs, the results given by SG13-MGKS fit well with the analytical solution  at different Prandtl numbers. The above results indicate that SG13-MGKS is able to correctly recover the results at different Prandtl numbers, thus validating the accuracy of the Shakhov collision mode in the present solver.

\subsection{\emph{Flow passing through NACA0012 airfoil}}
\label{sec3-2}

\begin{figure}[H]
\centering
\includegraphics[width=7cm]{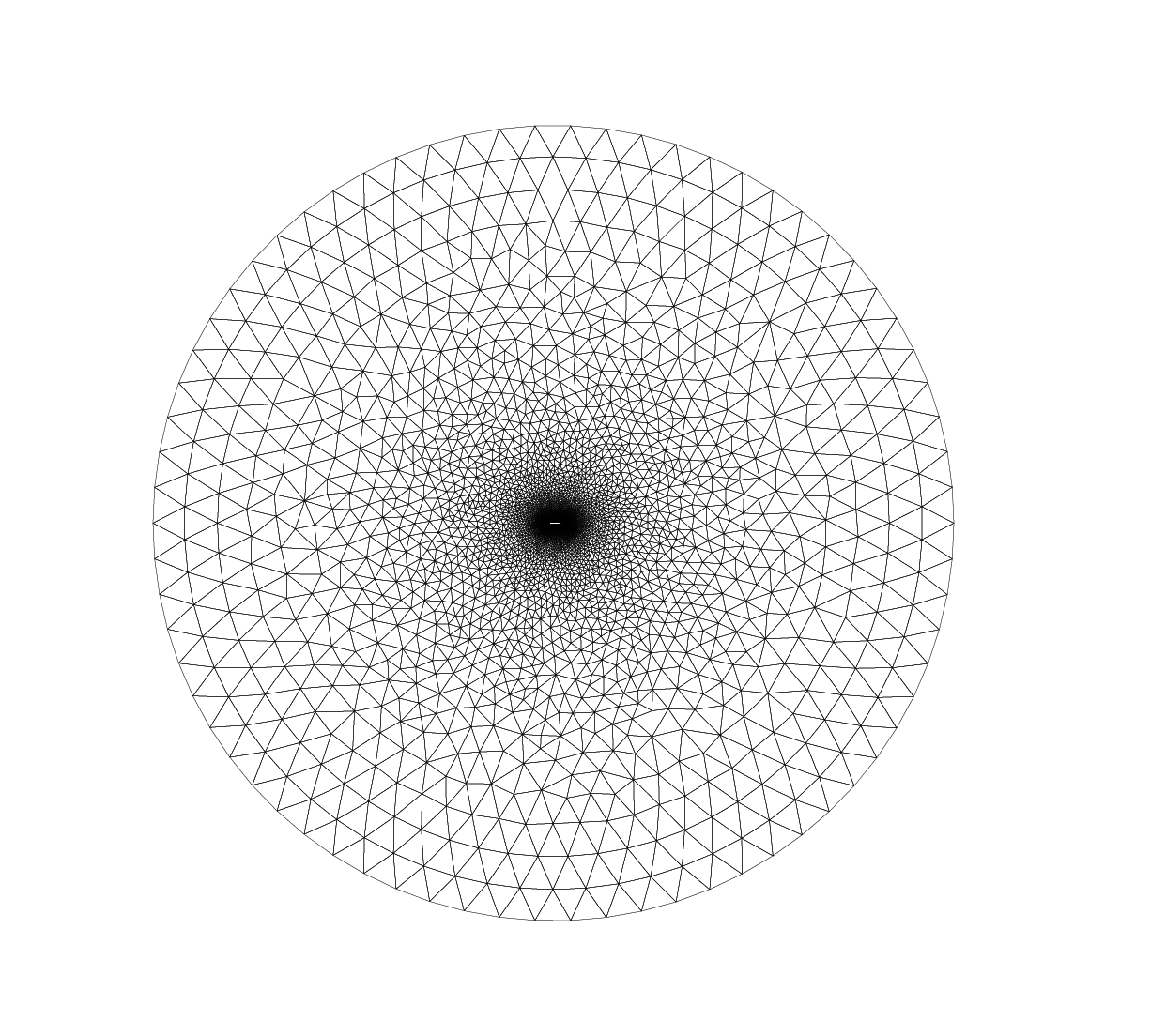}
\includegraphics[width=7cm]{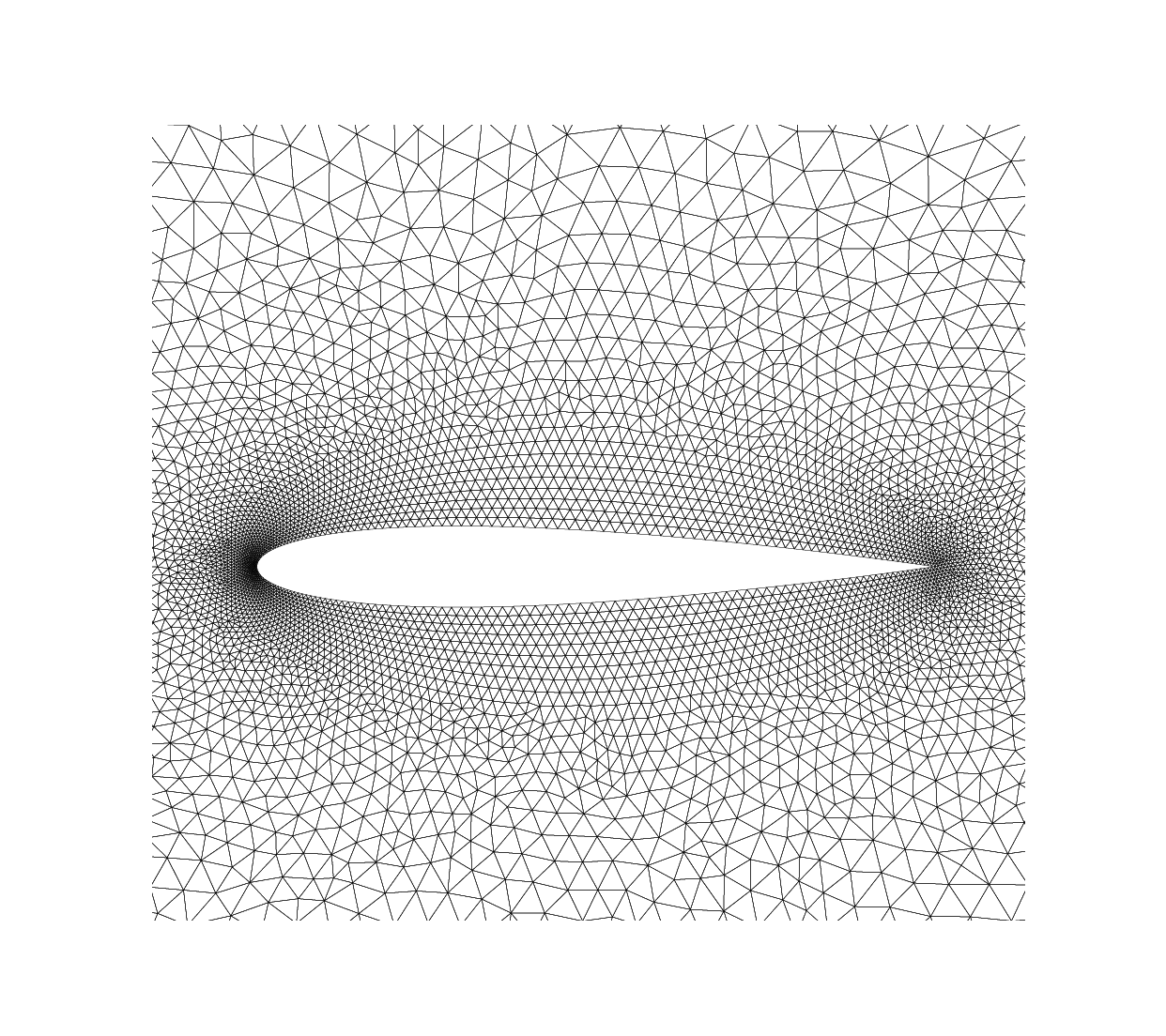}
\caption{Flow passing through a NACA0012 airfoil, (left) global mesh, (b) local enlargement of mesh.}
\label{Fig4}
\end{figure}

In the present case, the solver of SG13-MGKS is applied to simulate the flow past a NACA0012 airfoil where a triangular mesh with 16042 cells is utilized. The length of the airfoil is 1.0 and 201 mesh points are adopted on the airfoil. The diameter of the outer boundary is 50, and 71 mesh points are set on the outer boundary. The simulations involve the flow of argon gas with a Prandtl number of $\operatorname{Pr}=2/3$ and a specific heat ratio of $\gamma=5/3$. Fig. \ref{Fig4} presents the computing mesh and the localized enlargement close to the airfoil neighborhood.

\begin{figure}[H]
\centering
\includegraphics[width=7cm]{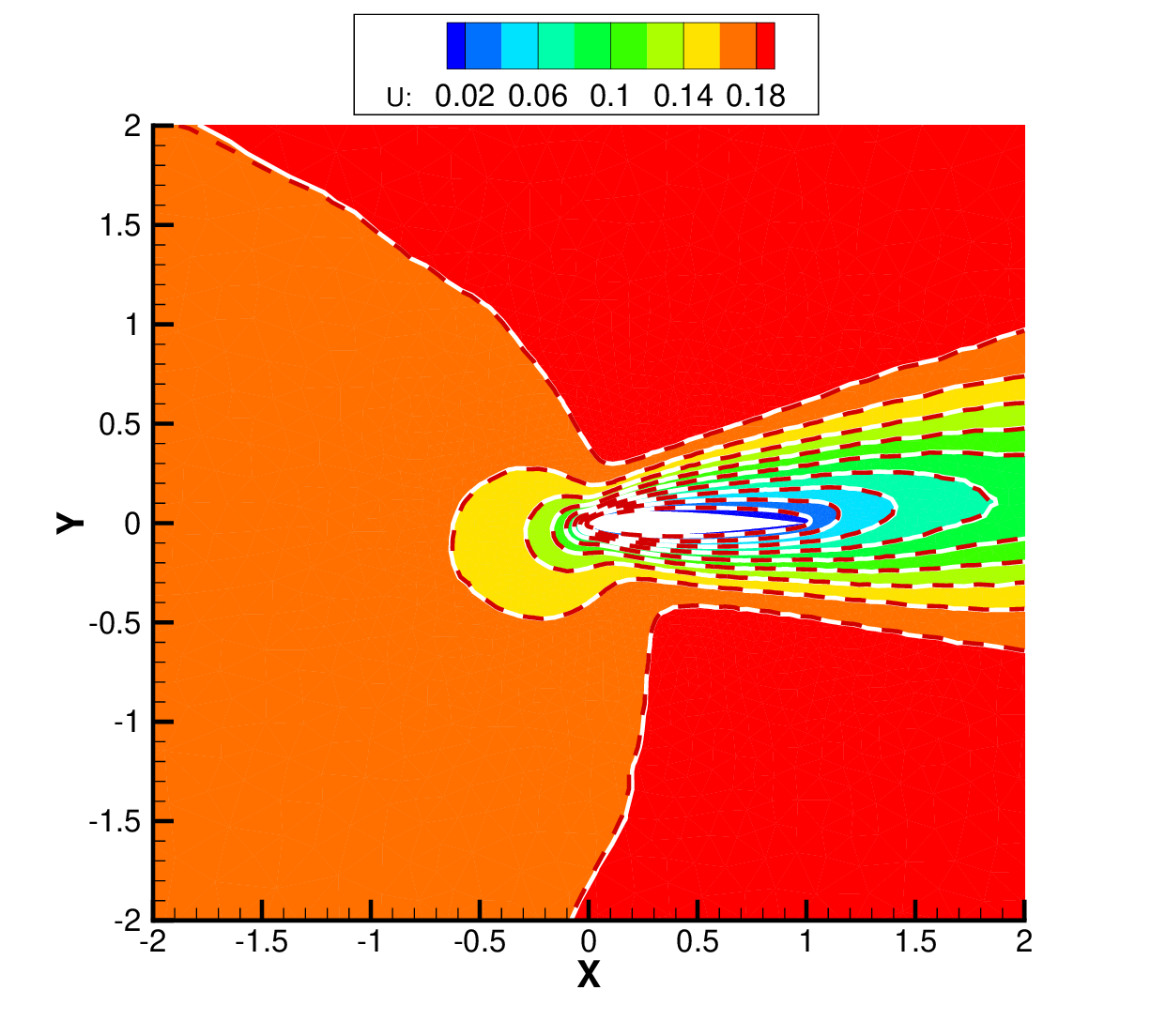}
\includegraphics[width=7cm]{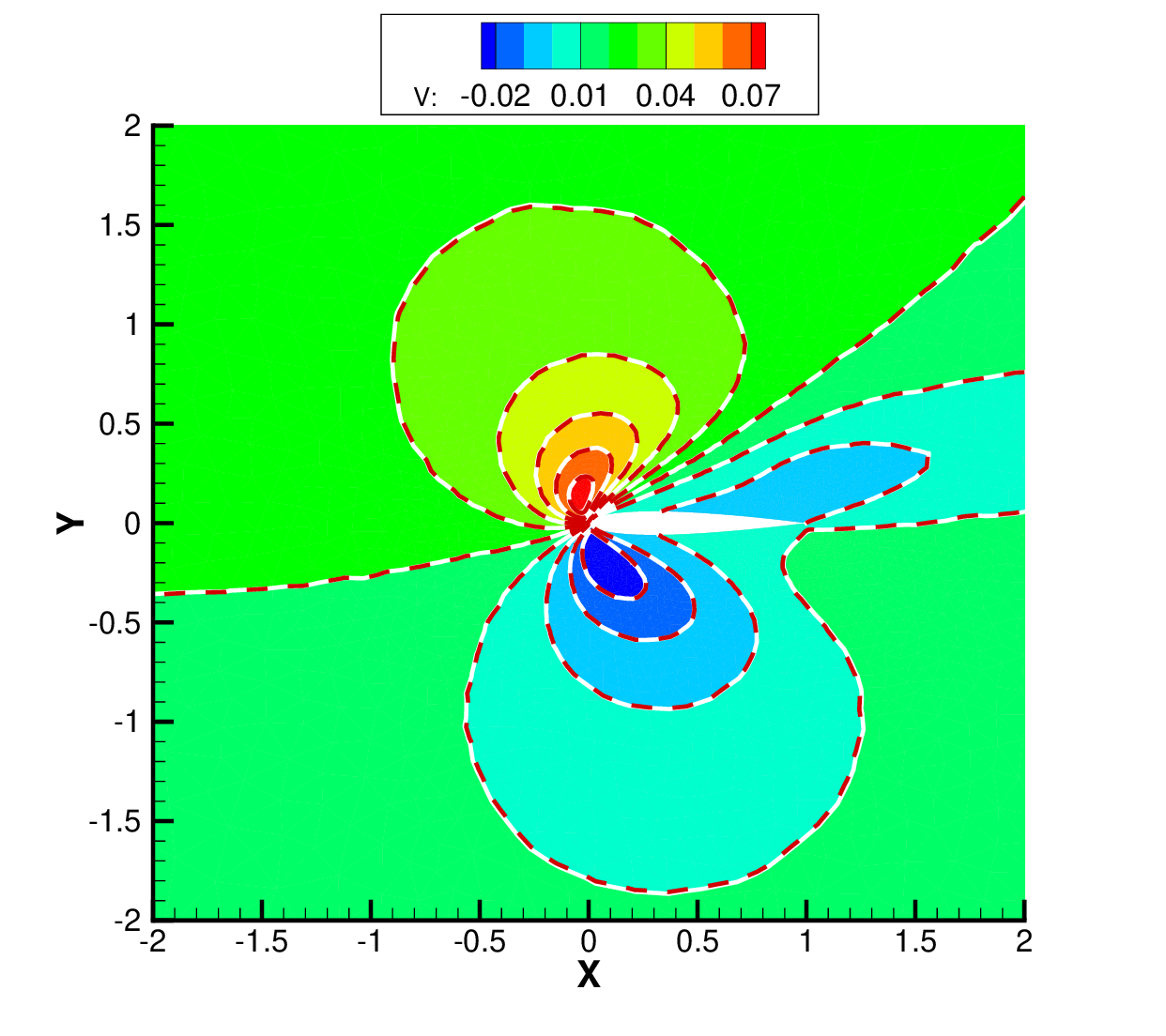}
\caption{Flow passing through a NACA0012 airfoil at $\mathrm{Ma}=0.2$ and $\mathrm{Kn}=0.01$, (left) $U$-velocity, (b) $V$-velocity. (Contour with white solid line: SG13-MGKS; Red dash line: IDVM)}
\label{Fig5}
\end{figure}

\begin{figure}[H]
\centering
\includegraphics[width=7cm]{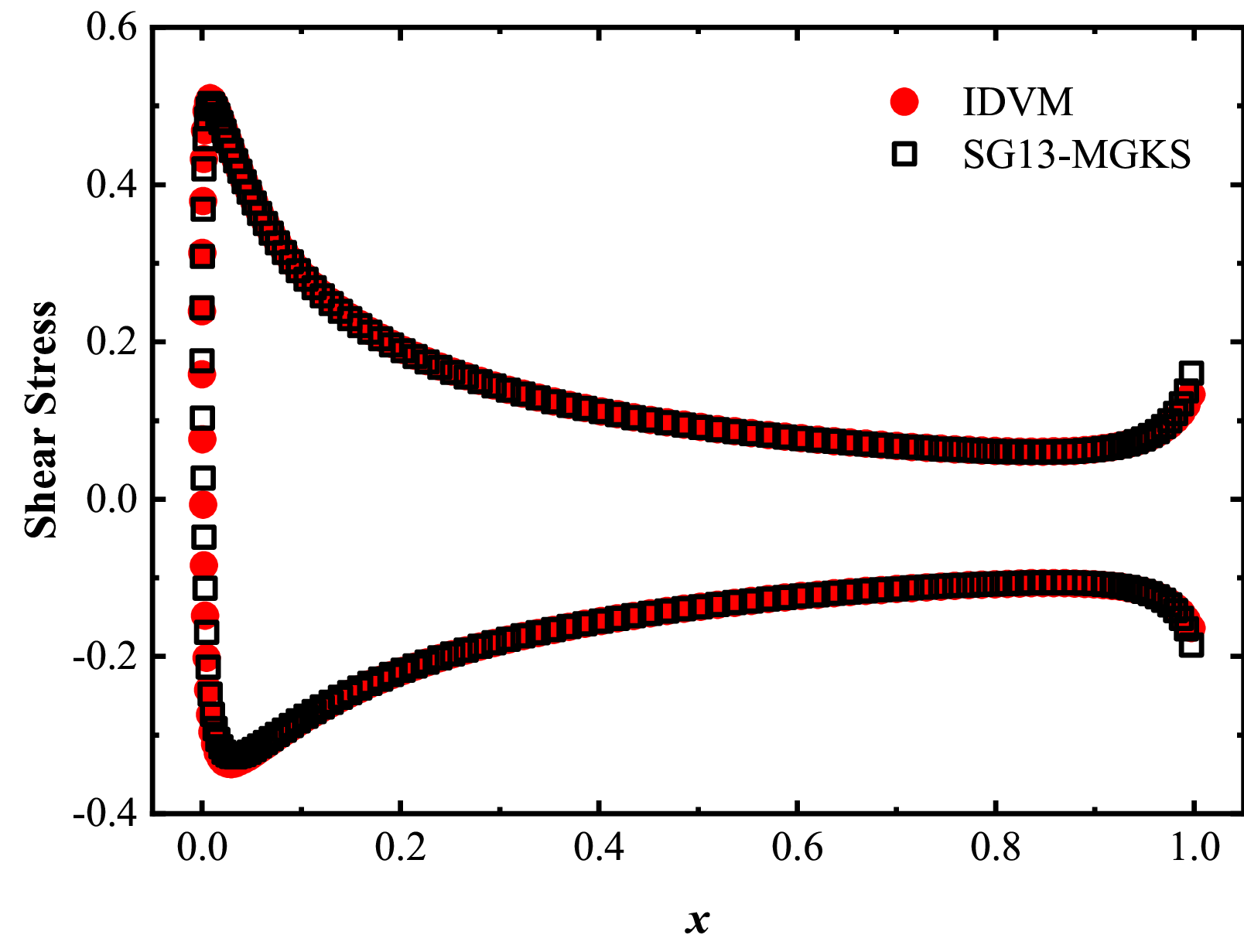}
\includegraphics[width=7cm]{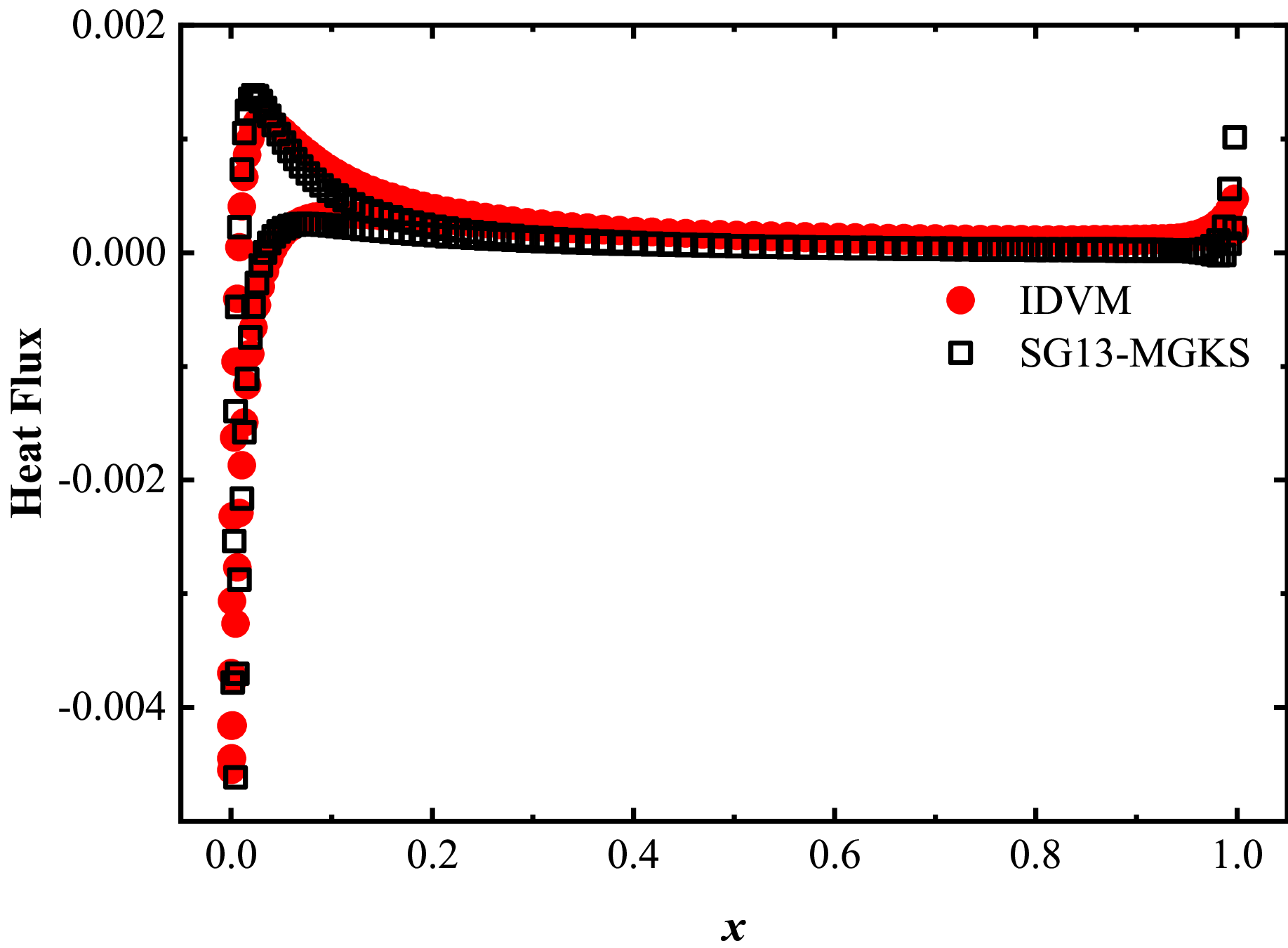}
\caption{Comparison of the distributions of (a) stress and (b) heat flux along the airfoil surface obtianed by the IDVM and the SG13-MGKS for flow around a NACA0012 airfoil at $\mathrm{Ma}=0.2$ and $\mathrm{Kn}=0.01$.}
\label{Fig6}
\end{figure}

In Fig. \ref{Fig5}, the comparsion of results for flow over the NACA0012 airfoil is presented. The reference Mach number and Knudsen number are set to 0.2 and 0.01, respectively. The colored contours with white solid lines represent $U$-velocity and $V$-velocity distributions obtained using the SG13-MGKS method. To verify the SG13-MGKS by the accurate result, the results from IDVM is chosen as the reference solution where the IDVM has been extensively validated in prior research \cite{yang_improved_2018, yang_implicit_2018} and widely acknowledged as an accurate method for the multiscale rarefied flows. As depicted in Fig. \ref{Fig5}, the results of $U$-velocity and $V$-velocity obtained by the SG13-MGKS method exhibit excellent agreement with the reference solution near the airfoil. Additionally, the distribution of shear stress and heat flux on the airfoil surface are presented in Fig. \ref{Fig6}. The results indicate that the solutions by SG13-MGKS closely match those results from the IDVM, with only small discrepancies observed in the edges of the airfoil. 

\begin{figure}[H]
\centering
\includegraphics[width=7cm]{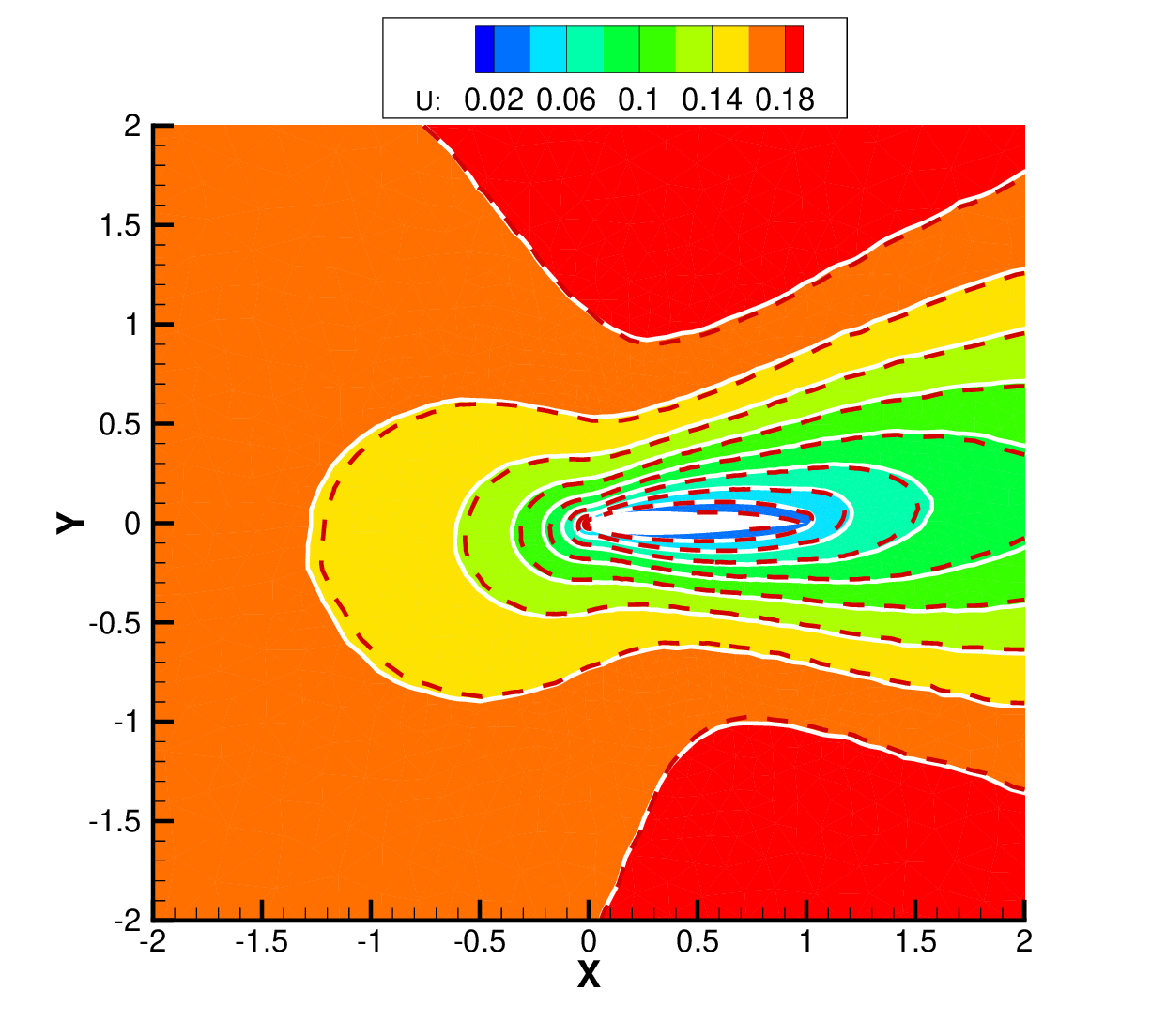}
\includegraphics[width=7cm]{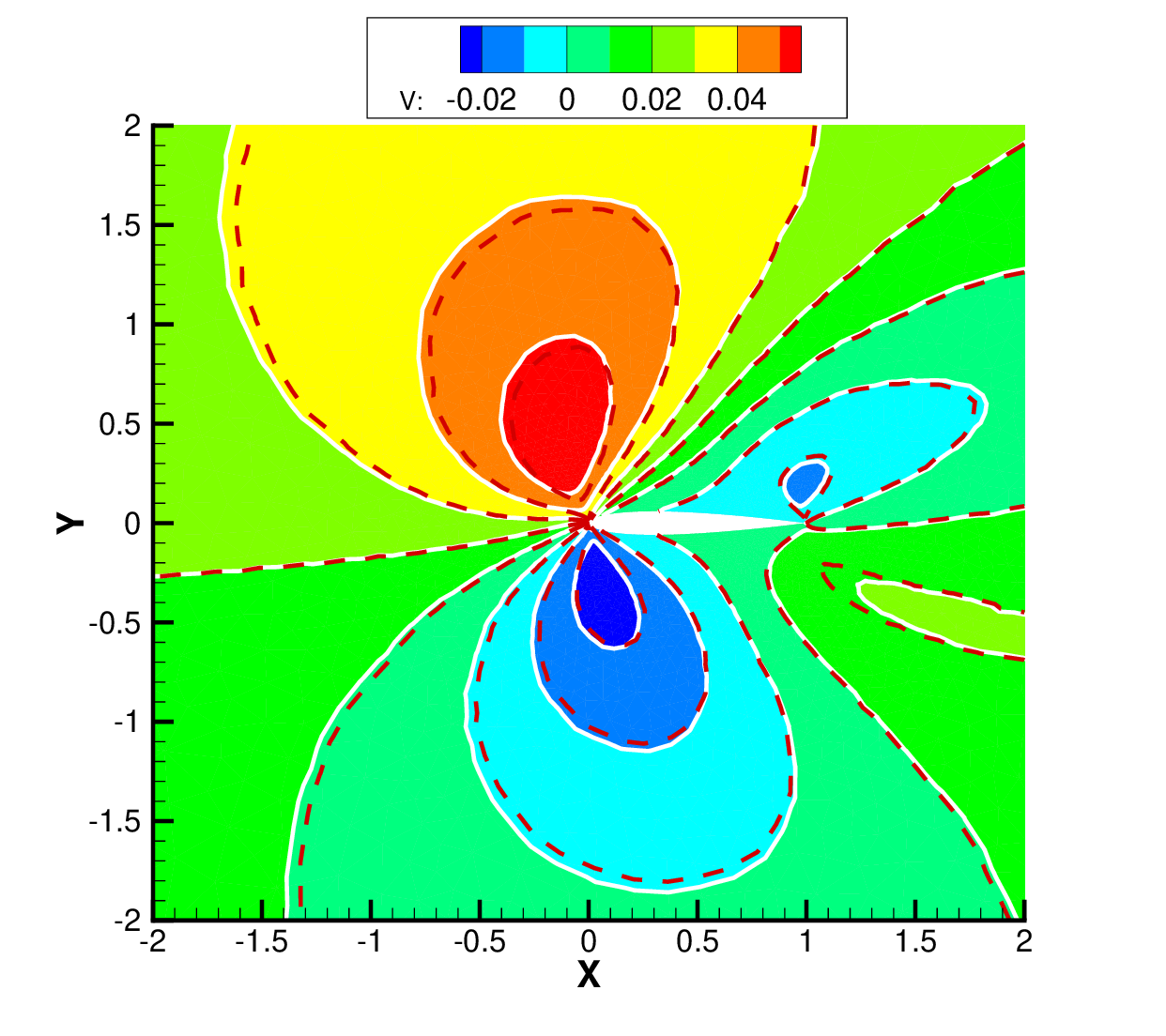}
\caption{Flow passing through a NACA0012 airfoil at $\mathrm{Ma}=0.2$ and $\mathrm{Kn}=0.05$, (left) $U$-velocity, (b) $V$-velocity. (Contour with white solid line: SG13-MGKS; Red dash line: IDVM)}
\label{Fig7}
\end{figure}

\begin{figure}[H]
\centering
\includegraphics[width=7cm]{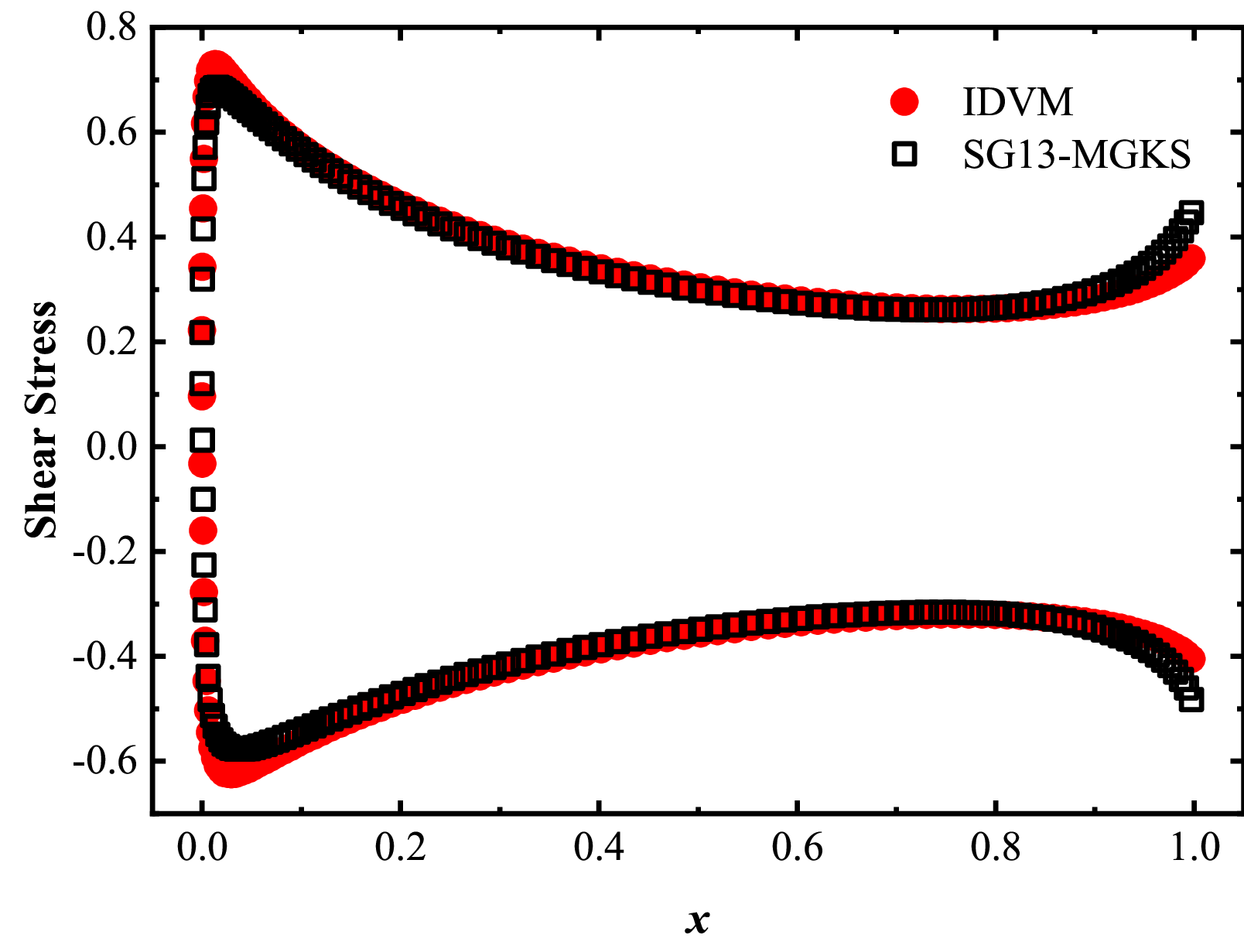}
\includegraphics[width=7cm]{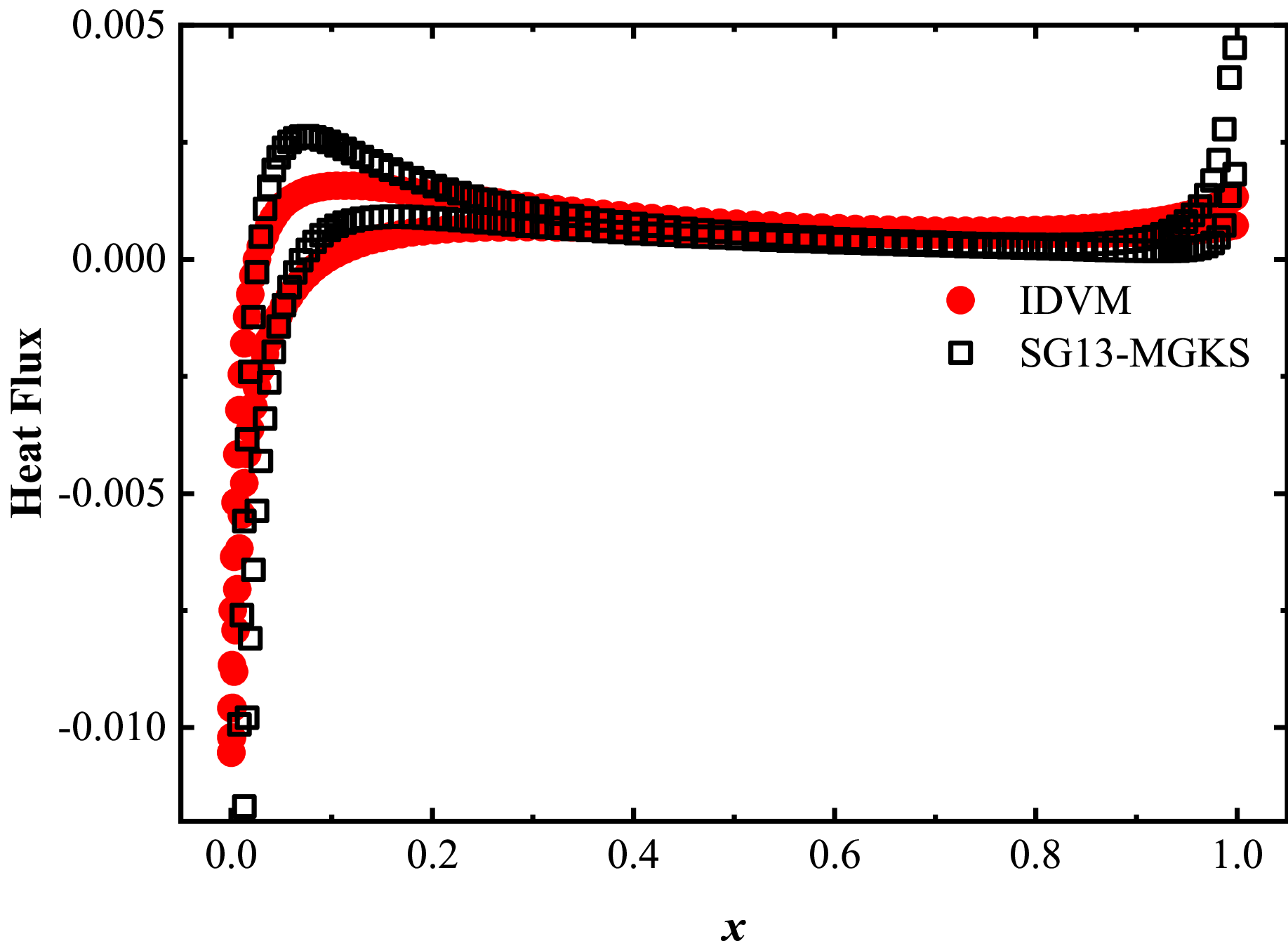}
\caption{Comparison of the distributions of (a) stress and (b) heat flux along the airfoil surface obtianed by the IDVM and the SG13-MGKS for flow around a NACA0012 airfoil at $\mathrm{Ma}=0.2$ and $\mathrm{Kn}=0.05$.}
\label{Fig8}
\end{figure}

When we increase the Knudsen number to 0.05, Fig. \ref{Fig7} presents thr illustration of the velocity field obtained by the approaches of SG13-MGKS and IDVM. It is observed that SG13-MGKS still can provide results consistent with reference data at Knudsen number equals 0.05, with only minor discrepancies evident. In addition, the distribution of stress and heat flux presented in Fig. \ref{Fig8} reveal that SG13-MGKS performs better in predicting stress compared to heat flux. This phenomenon may be attributed to two potential factors. Firstly, heat flux involves higher-order moments compared to stress, making it potentially more challenging to approximate by using the G13 distribution function. Another reason could be that, due to the near-isothermal nature of the flow field, the numerical values of heat flux are smaller, making errors more apparent. Overall, SG13-MGKS manages to provide satisfactory computation results.

\begin{figure}[H]
\centering
\includegraphics[width=7cm]{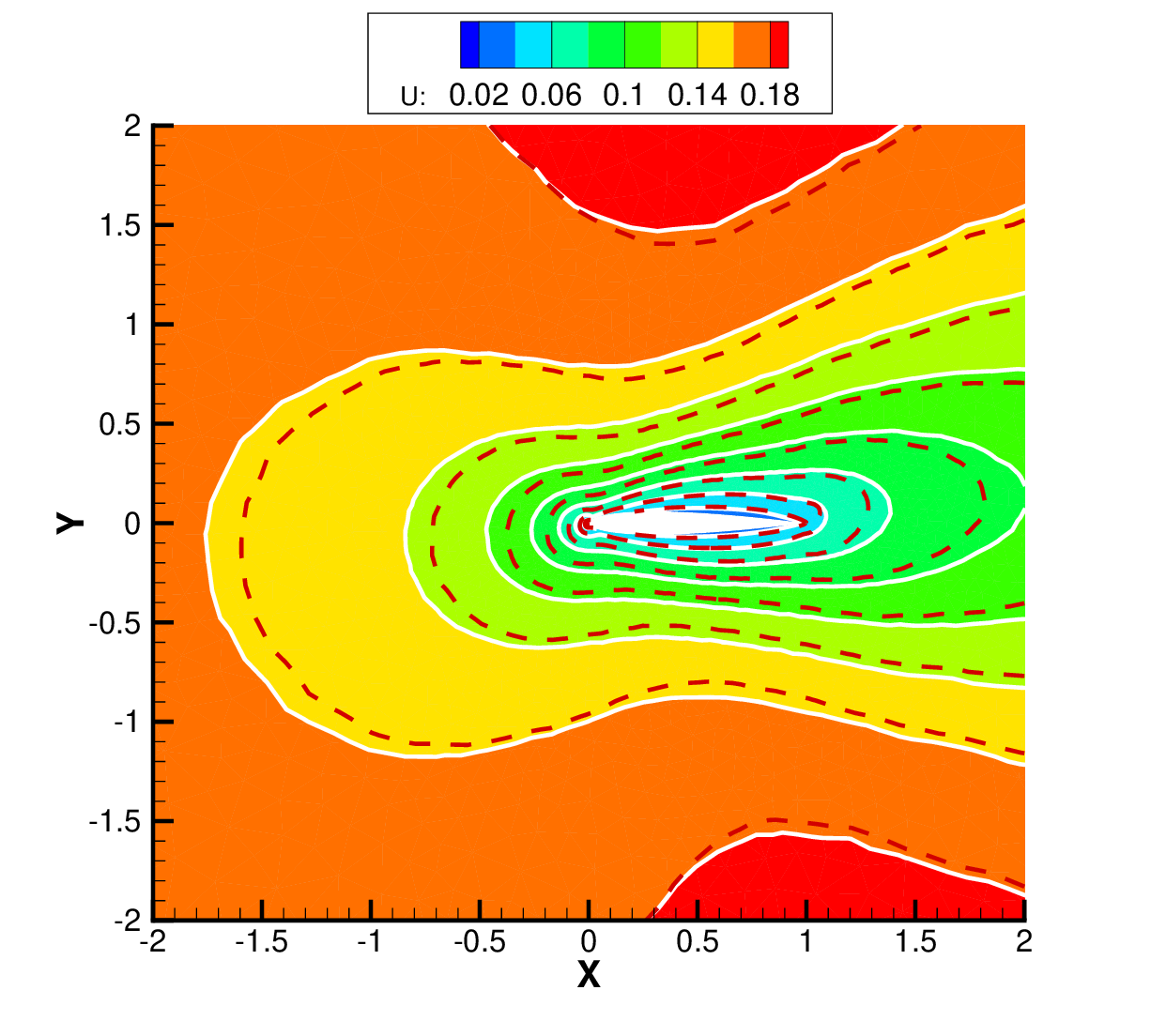}
\includegraphics[width=7cm]{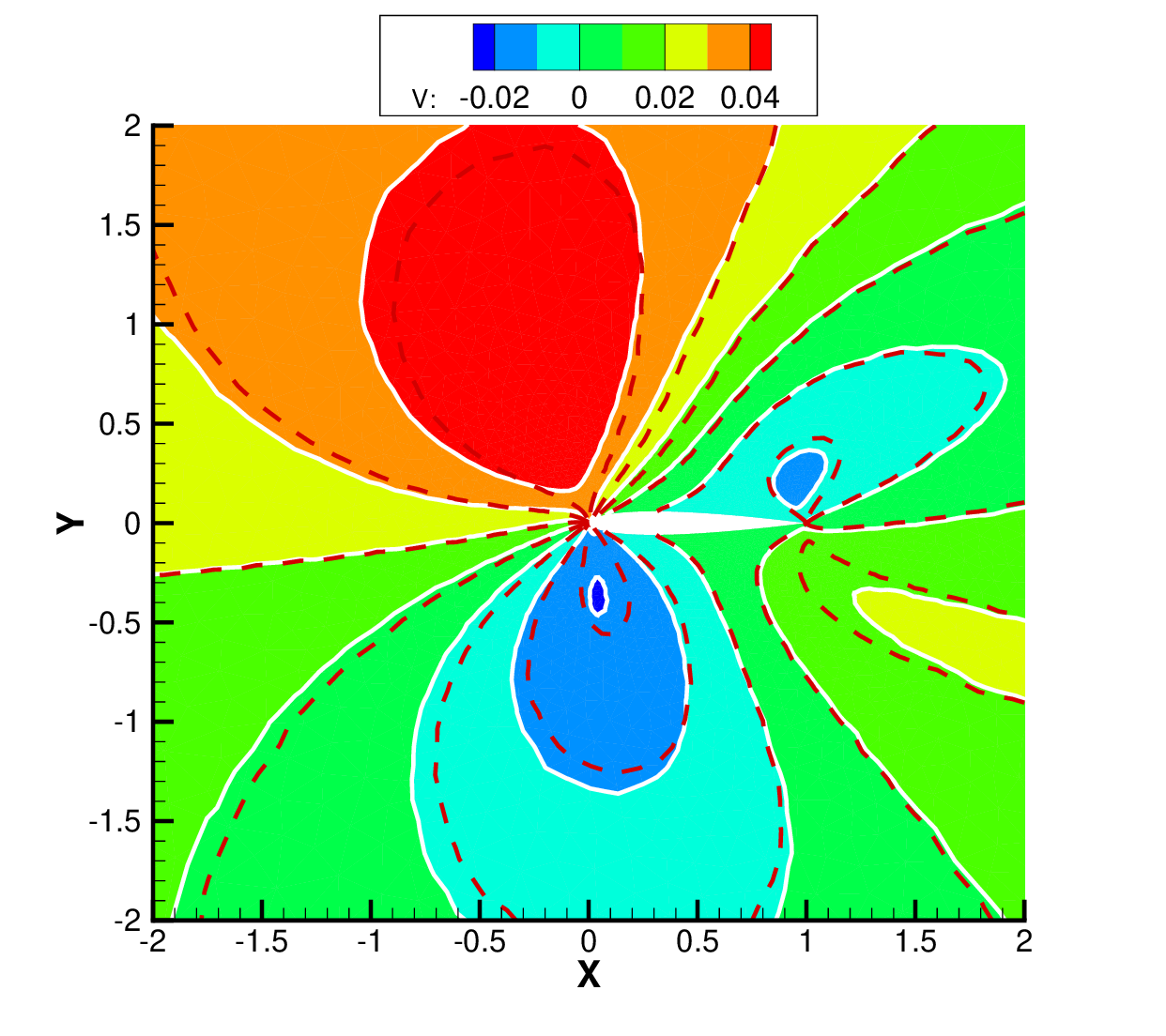}
\caption{Flow passing through a NACA0012 airfoil at $\mathrm{Ma}=0.2$ and $\mathrm{Kn}=0.1$, (left) $U$-velocity, (b) $V$-velocity. (Contour and white solid line: SG13-MGKS; Red dash line: IDVM)}
\label{Fig9}
\end{figure}

\begin{figure}[H]
\centering
\includegraphics[width=7cm]{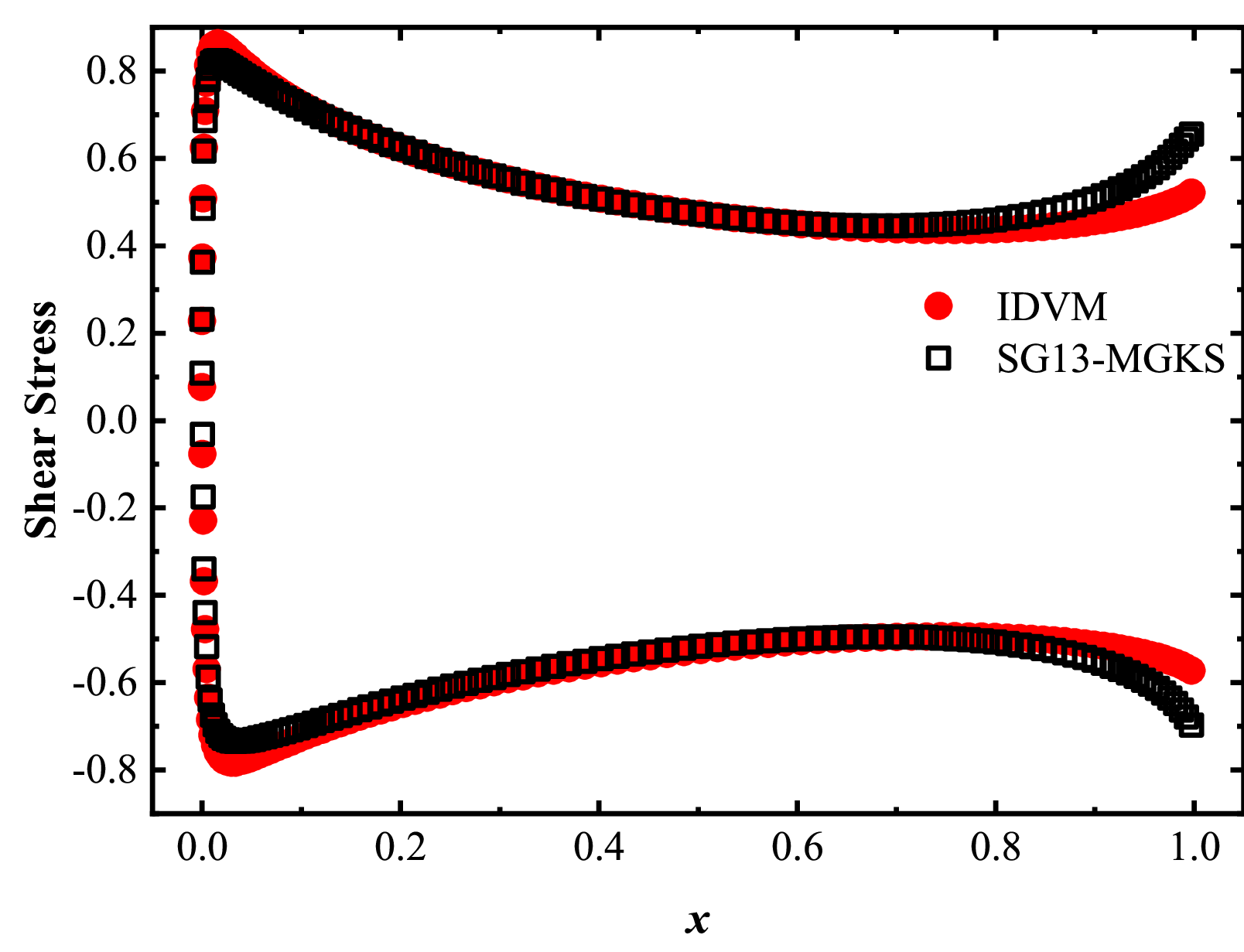}
\includegraphics[width=7cm]{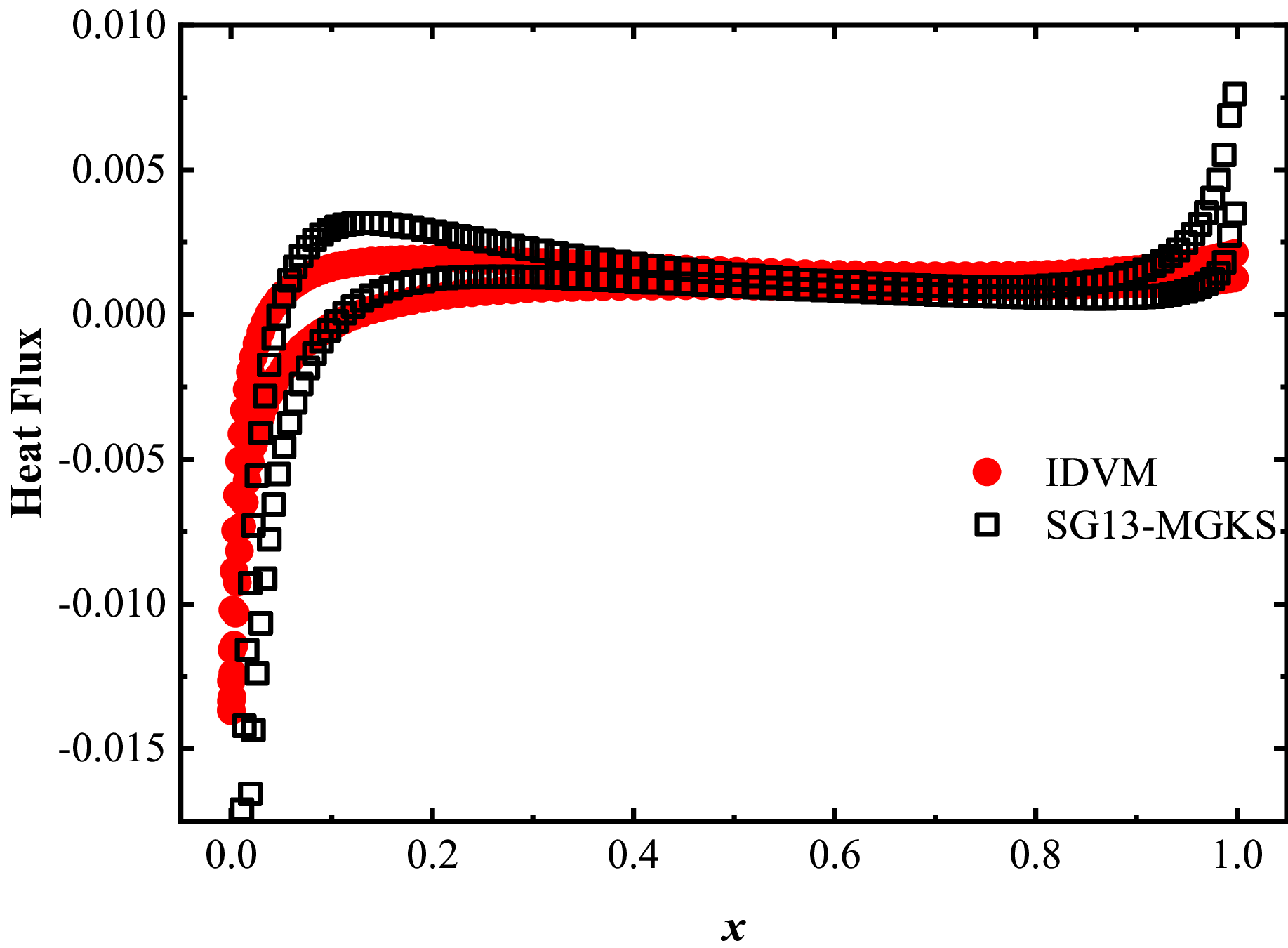}
\caption{Comparison of the distributions of (a) stress and (b) heat flux along the airfoil surface obtianed by the IDVM and the SG13-MGKS for flow around a NACA0012 airfoil at $\mathrm{Ma}=0.2$ and $\mathrm{Kn}=0.1$.}
\label{Fig10}
\end{figure}

Further increase the Knudsen number to 0.1, the results of velocity contours presented in Fig. \ref{Fig9} indicating that the disturbance of the airfoil on the flow field become expanding due to the increased rarefied effect. This phenomenon reflects the interaction between the wall and the gas flow becoming more dominant due to the growing significance of rarefaction effects. In this scenario, SG13-MGKS continues to provide reasonable predictions, albeit with some deviations relative to the reference results. The distributions of stress and heat flux shown in  Fig. \ref{Fig10} reveal that notable discrepancies are primarily concentrated at the leading and trailing edges of the airfoil. 

In Table \ref{tb1}, we have presented predictions of lift coefficient and drag coefficient, along with relative errors, obtained using various methods. The results indicate that SG13-MGKS outperforms predictions for the lift coefficient across different Knudsen numbers, as compared to predictions for the drag coefficient. Furthermore, the relative errors increase at the higher Knudsen numbers, while the overall error can be controlled under 5\% at Knudsen number below 0.1,. 

Besides, the computational times and memory consumption of IDVM, original version of G13-MGKS and Simplified G13-MGKS are given in Table \ref{tb2}. The IDVM is adopted the Gauss-Hermite quadrature with $28 \times 28$ points in the velocity space and all three methods use the same physical mesh. By eliminating the discretization of the velocity space, both G13-MGKS and SG13-MGKS can significantly reduce computation time and memory consumption compared to IDVM. It is worth noting that the simplified version of G13-MGKS shows a better efficiency advantage over the original version of G13-MGKS, thanks to that our simplified treatment of numercial fluxes enables the number of gradients that need to be computed in SG13-MGKS to be greatly reduced. Considering that aforementioned results provide support that SG13-MGKS can be a accurate computational method for the flow problems at moderate Knudsen number, It demonstrates that THE balance between efficiency and accuracy can be achieved by the SG13-MGKS with the applicability for the unstructured mesh.

\begin{table}[H]
\centering
\setlength{\abovecaptionskip}{10pt}
\setlength{\belowcaptionskip }{10pt}
\caption{Drag coefficient (Cd) and Lift coefficient (CL) for flow past a NACA0012 airfoil}
\setlength{\tabcolsep}{9mm}
\begin{threeparttable}
\begin{tabular}{cccc}
\hline
\hline
\textbf{Cd/CL}& \textbf{Kn = 0.01} & \textbf{Kn = 0.05}& \textbf{Kn = 0.1} \\
\hline
IDVM    & 0.73/0.41&   1.79/0.56&   2.54/0.70\\
SG13-MGKS    & 0.73/0.41&   1.84/0.57&   2.66/0.73   \\
Relative Error    & 0.0\%/0.0\%&   2.79\%/1.78\%&   4.72\%/4.28\%   \\
\hline
\hline
\end{tabular}
\label{tb1}
\end{threeparttable}       
\end{table}

\begin{table}[H]
\centering
\setlength{\abovecaptionskip}{10pt}
\setlength{\belowcaptionskip }{10pt}
\caption{Computational times and memory consumption of different methods for the flow around a NACA0012 airfoil.}
\setlength{\tabcolsep}{7mm}
\begin{threeparttable}
\begin{tabular}{ccccc}
\hline
\hline
\textbf{}& \textbf{IDVM}& \textbf{G13-MGKS}& \textbf{SG13-MGKS} & \textbf{SR$^{\dagger}$}\\
\hline
\textbf{Time}& 37.95 hours&   0.87 hours&   0.22 hours&   172.2\\
\textbf{Memory}& 1638 MB&   262 MB&   83 MB&   19.8\\
\hline
\hline
\end{tabular}
\begin{tablenotes}    
\footnotesize              
\item[$^{\dagger}$] Speed-up Ratio (SR) between the SG13-MGKS and IDVM in terms of computational time 
\end{tablenotes}   

\label{tb2}
\end{threeparttable}
\end{table}

\subsection{\emph{Pressure-driven flow in a variable-diameter circular pipe}}
\label{sec3-3}

\begin{figure}[H]
\centering
\includegraphics[width=10cm]{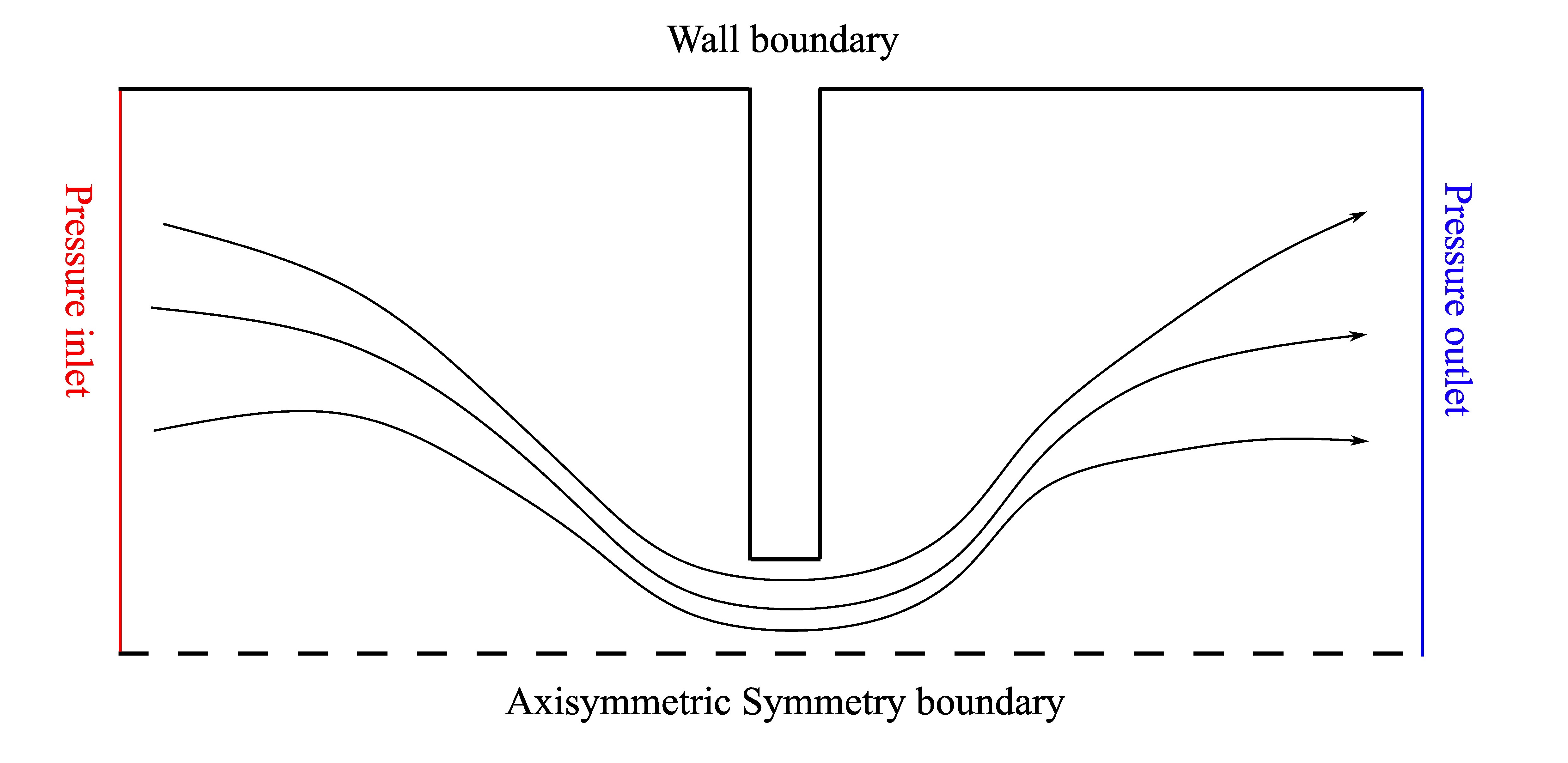}
\caption{Schematic of the pressure-driven micro-flow.}
\label{Fig11}
\end{figure}

\begin{figure}[H]
\centering
\includegraphics[width=9cm]{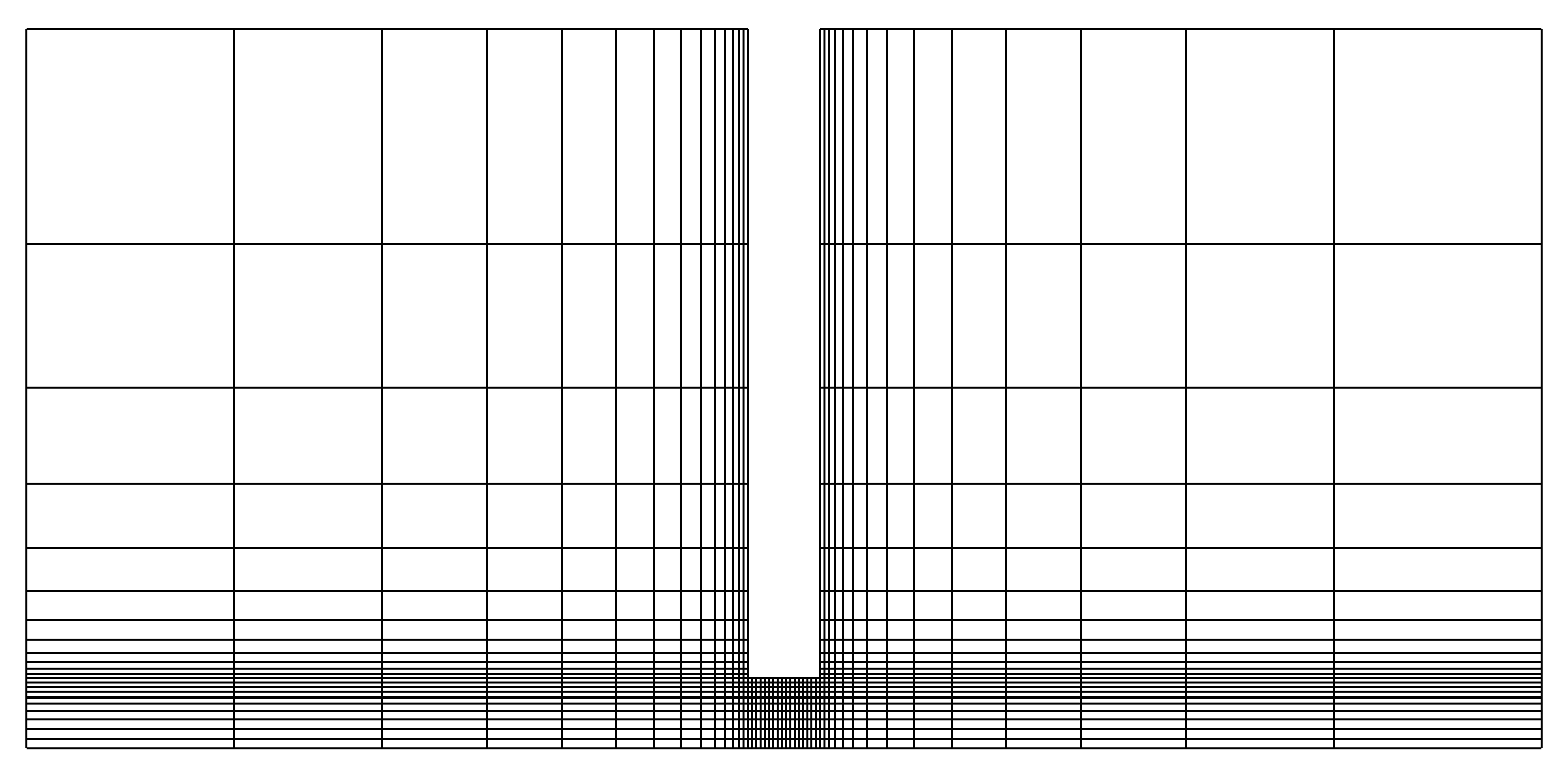}
\caption{Mesh of the pressure-driven micro-flow}
\label{Fig12}
\end{figure}

To further verify the SG13-MGKS in simulating on rarefied flows, the pressure-driven flow in a variable-diameter circular pipe is studied as the present case. The Schematic is shown in Fig. \ref{Fig11}, where a round pipe connect the two larger channels. The left and right boundaries of the computational region are set to pressure inlet and outlet conditions, respectively. The ratio of inlet pressure $p_{1}$ to outlet pressure $p_{2}$ is 2.0, and same temperature is applied to the inlet boundary, outlet boundary and the fixed wall as $T_{0} = 1.0$. In order to simulate the flow in a circular pipe, the lower side of the computational region is set to axisymmetric boundary condition. The height of the inlet and outlet channels is 10 and the height of the middle channel is 1. The width of inlet and outlet channels is 10 and the width of the middle channel is 1. The viscosity could be calculated by 

\begin{equation}
\mu=\mu_{r e f}\left(\frac{T}{T_{0}}\right)^{\omega}, 
\label{eq39}
\end{equation}

\noindent where the temperature dependency index is adopted as the $\omega=0.5$ for the hard-sphere (HS) model. The 874 hexahedral cells are adopted to discretize the computational region by using three level meshes, which are shown in Fig. \ref{Fig12}.  

\begin{figure}[H]
\centering
\subfigure[]{
\includegraphics[width=7cm]{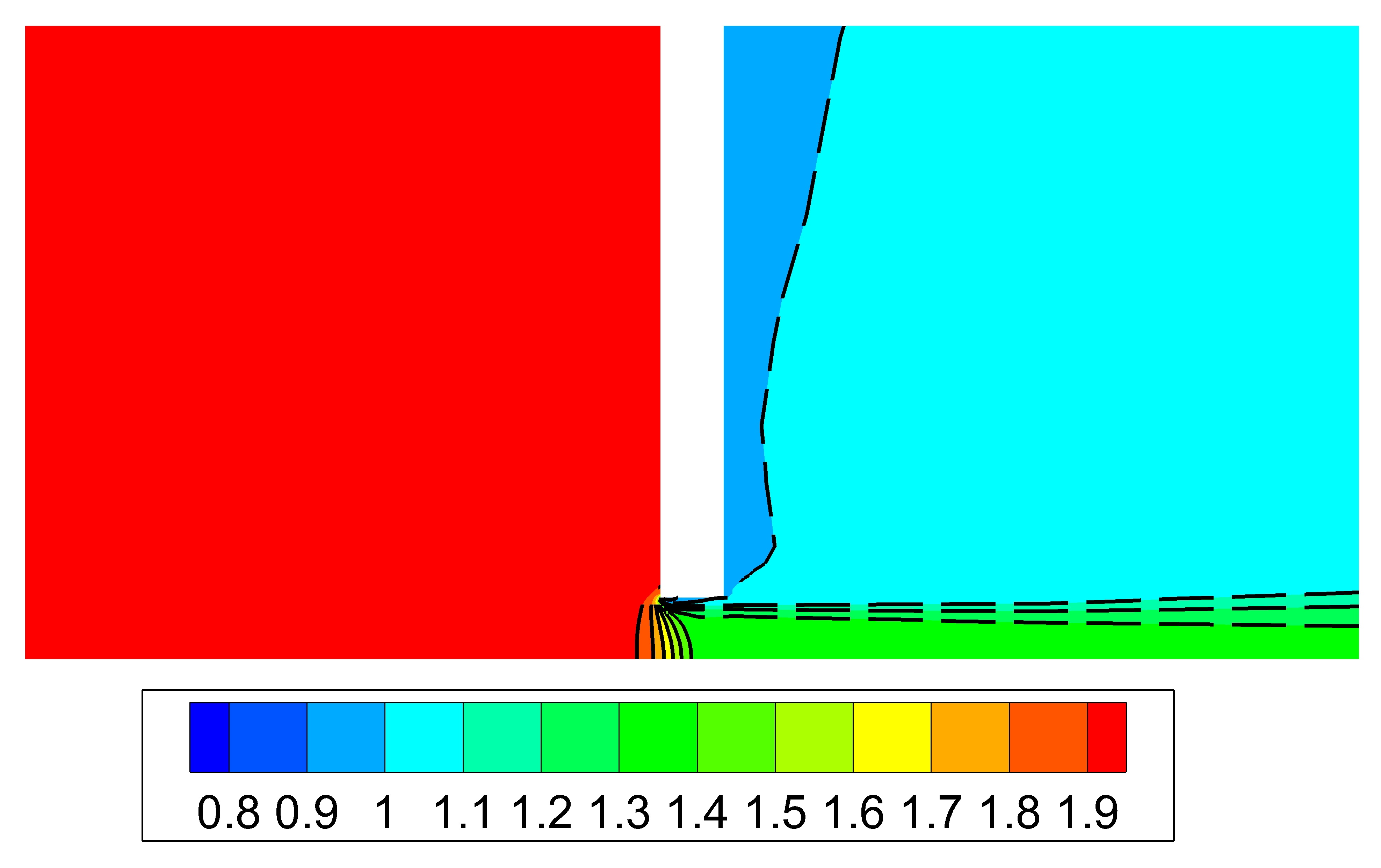}
}
\quad
\subfigure[]{
\includegraphics[width=7cm]{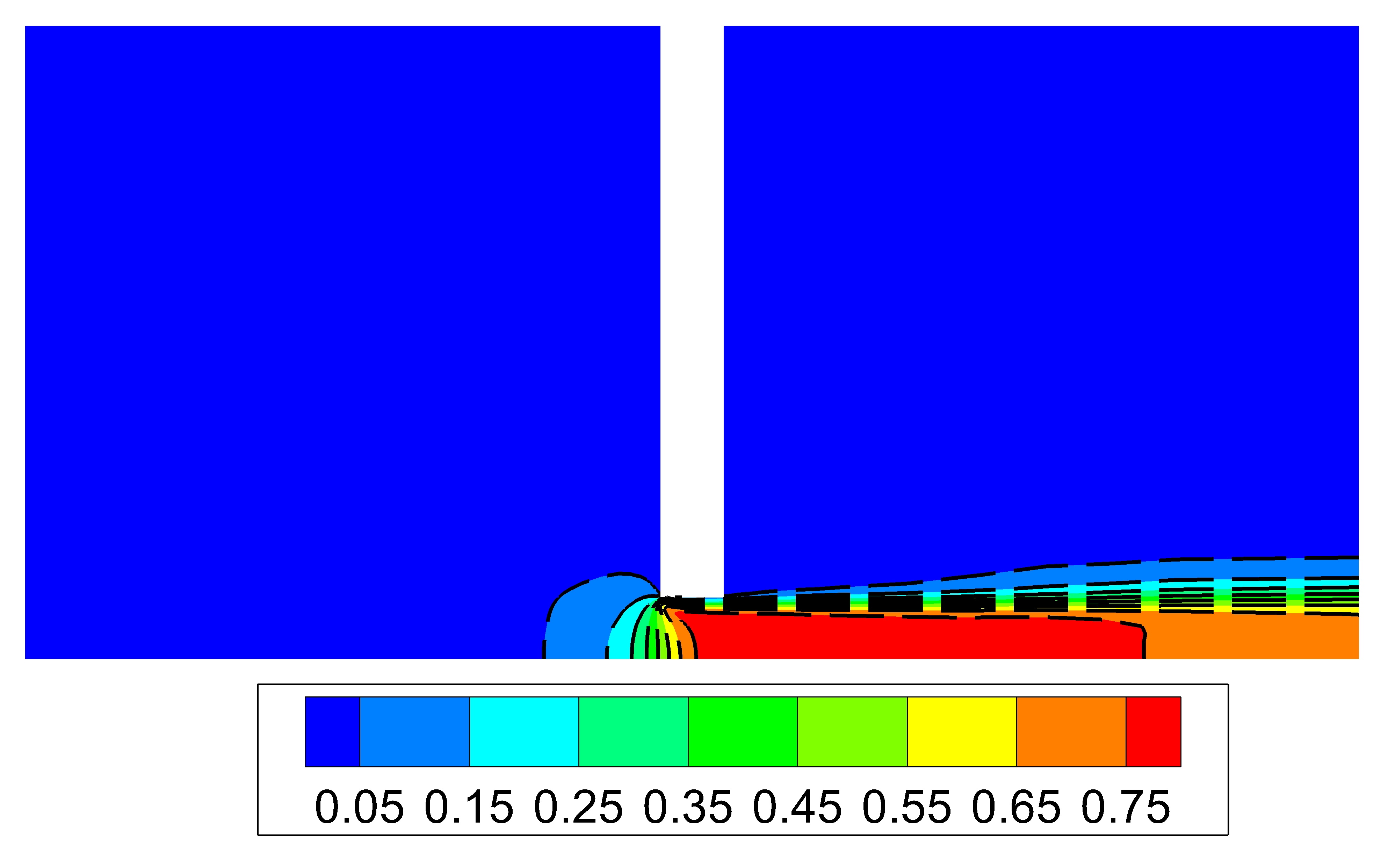}
}
\caption{Comparison of contours in the pressure-driven flow passing through a pipe at $\mathrm{Kn}=0.00181$, (a) density and (b) $U$-velocity. (Contour: SG13-MGKS; Black dash line: IDVM)}
\label{Fig13}
\end{figure}

\begin{figure}[H]
\centering
\subfigure[]{
\includegraphics[width=7cm]{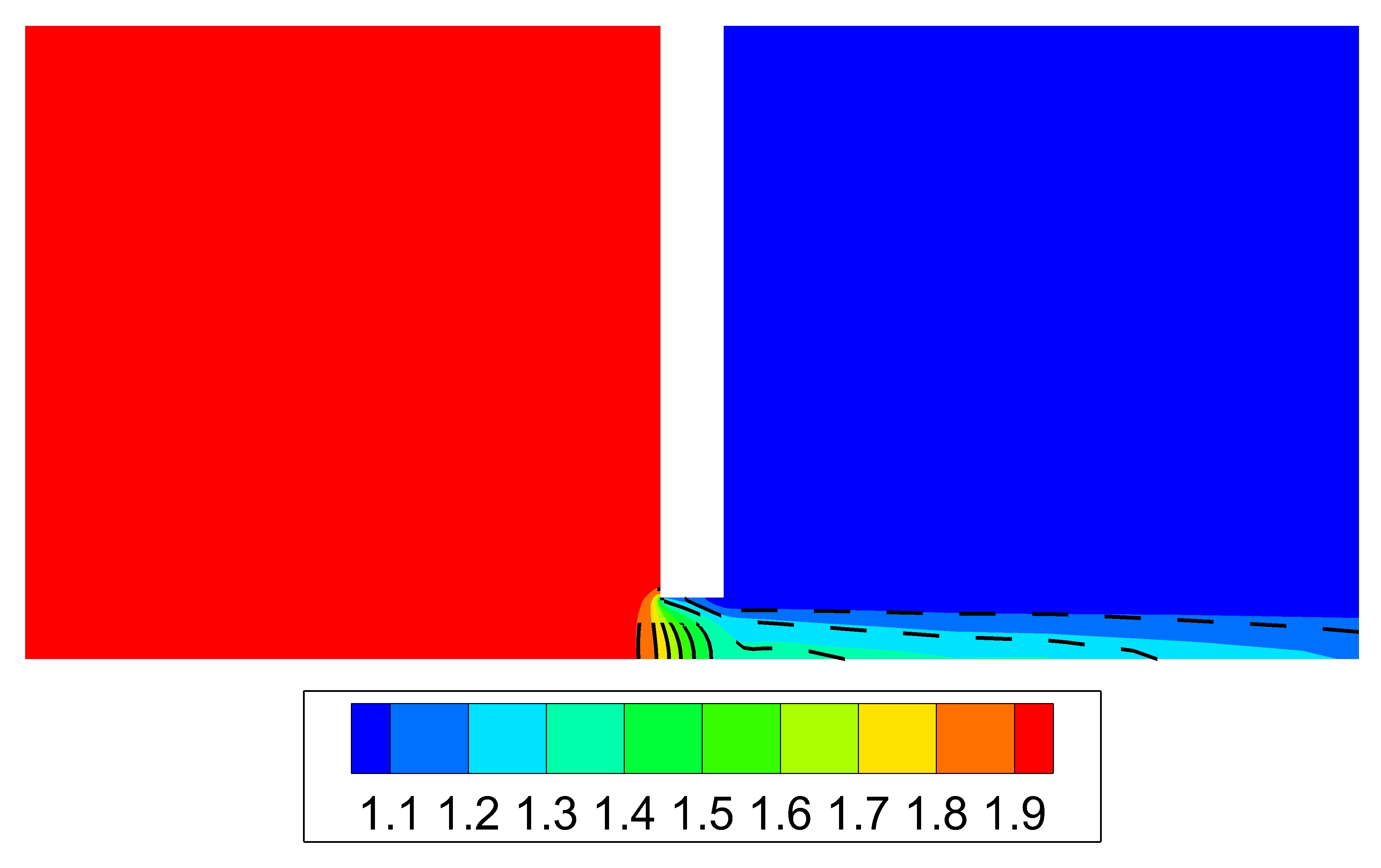}
}
\quad
\subfigure[]{
\includegraphics[width=7cm]{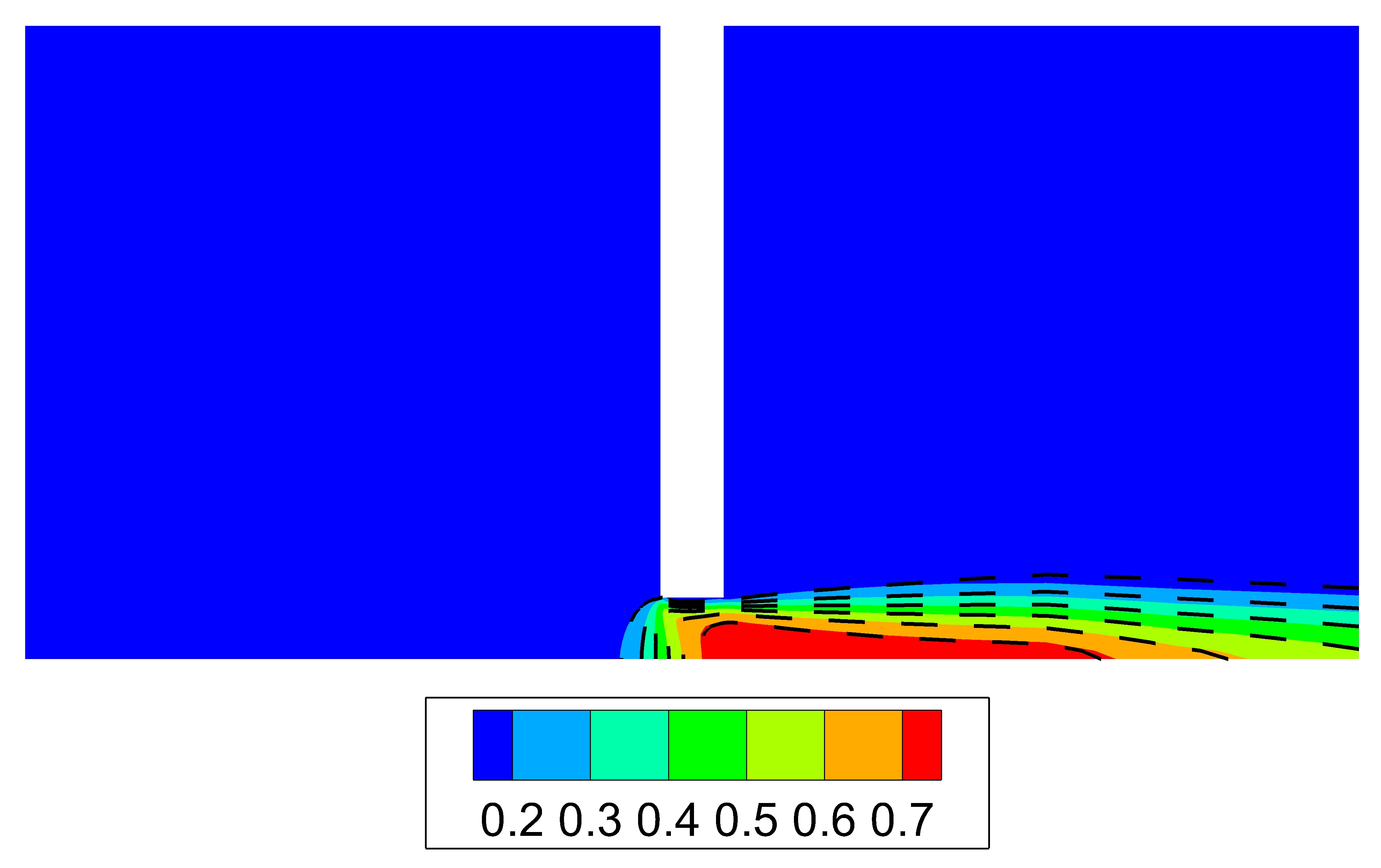}
}
\caption{Comparison of contours in the pressure-driven flow passing through a pipe at $\mathrm{Kn}=0.0181$, (a) density and (b) $U$-velocity. (Contour: SG13-MGKS; Black dash line: IDVM)}
\label{Fig14}
\end{figure}

\begin{figure}[H]
\centering
\subfigure[]{
\includegraphics[width=7cm]{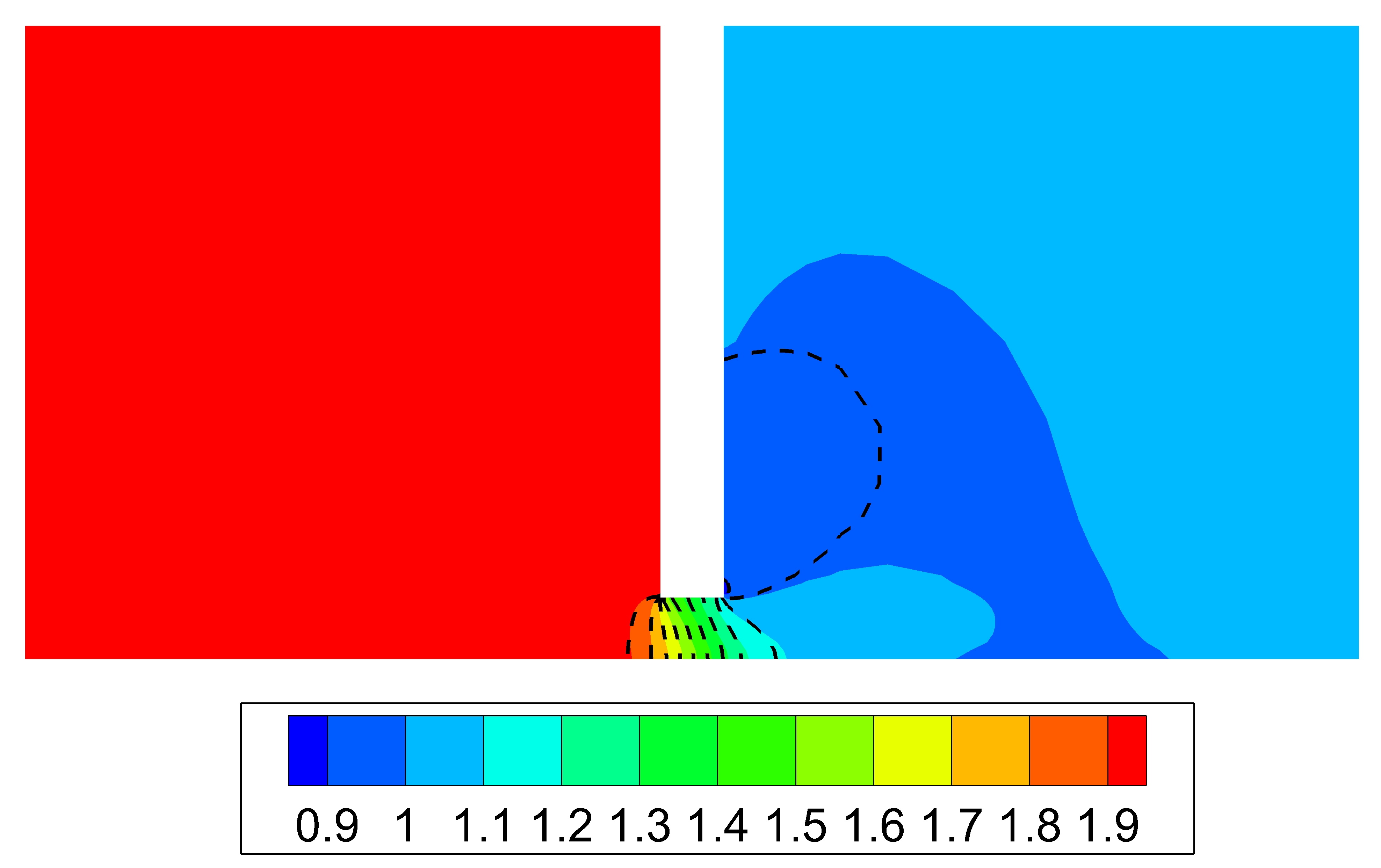}
}
\quad
\subfigure[]{
\includegraphics[width=7cm]{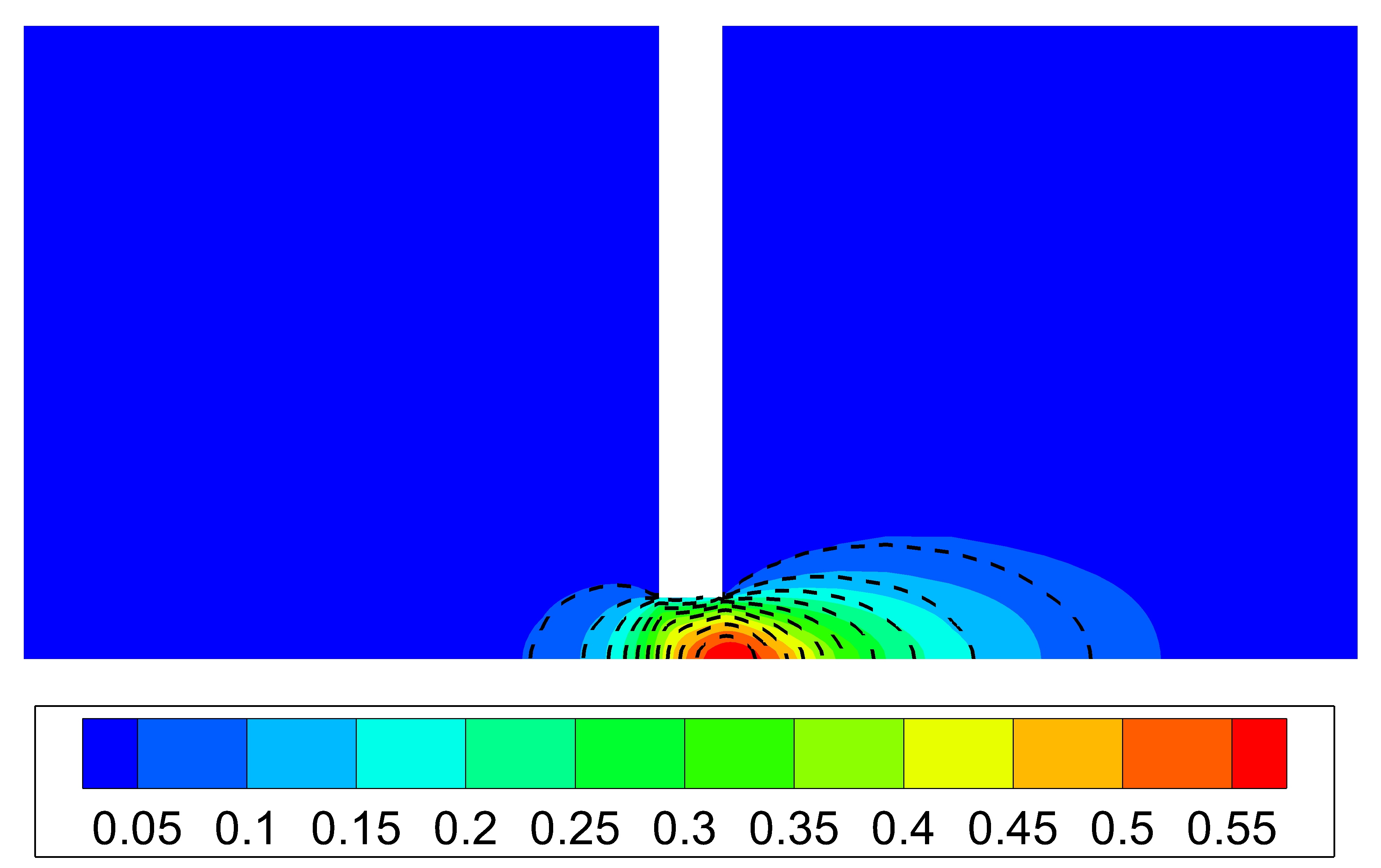}
}
\caption{Comparison of contours in the pressure-driven flow passing through a pipe at $\mathrm{Kn}=0.181$, (a) density and (b) $U$-velocity. (Contour: SG13-MGKS; Black dash line: IDVM)}
\label{Fig15}
\end{figure}

To validate the accuracy of our current method under varying degrees of rarefaction, we present density and velocity contours obtained by using SG13-MGKS in Figs. (\ref{Fig13}) - (\ref{Fig15}). These plots correspond to Knudsen numbers of 0.00181, 0.0181, and 0.181, covering the transition from slip flow to transition flow regimes. Additionally, results computed via the IDVM are included in the figures as black dashed lines for the comparison. Our findings reveal that at a Knudsen number of 0.00181, SG13-MGKS provides results that closely match those obtained by IDVM, thereby validating the accuracy of SG13-MGKS in near-continuum flow regions. As the Knudsen number increases to 0.0181, very subtle differences between SG13-MGKS and the reference results become noticeable, but the current method still provides excellent predictions. However, as rarefaction effects intensify, reaching a Knudsen number of 0.181, more obvious discrepancies appear in the density and velocity contour plots. These discrepancies are primarily observed in the outlet region of the pipe, where localized rarefaction effects become more pronounced, resulting in deviations in the predictions based on the G13 distribution function.

In order to qualitatively illustrate the performance of SG13-MGKS in pressure-driven pipe flow, we present the results of reduced mass flow rate in Table \ref{tb3}, which are compared with those obtained through the Direct Simulation Monte Carlo (DSMC) method. The calculation formula for the reduced mass flow rate $\hat{M}$ is provided below

\begin{equation}
\hat{M}=\frac{\sqrt{2 \pi R_g T_0}}{p_1} \int_{A(x)} \rho(x, y) U(x, y) d y,
\label{eq40}
\end{equation}

Our comparison reveals that at a Knudsen number of 0.181, SG13-MGKS exhibits an error of approximately 1.62\%. Furthermore, the comparative analysis of computational efficiency and memory consumption presented in Table \ref{tb4} demonstrates that SG13-MGKS achieves computational efficiency nearly fifty times greater than that of IDVM and four times greater than that of orginal version of G13-MGKS, along with significantly reduced memory consumption. These results underscore the accuracy and efficiency of SG13-MGKS in computing the multiscale flow issues from continuum regime to moderately rarefied flow regime.

\begin{table}[H]
\centering
\setlength{\abovecaptionskip}{10pt}
\setlength{\belowcaptionskip }{10pt}
\caption{Mass flow rate for pressure-driven flow through a pipe.}
\setlength{\tabcolsep}{7mm}
\begin{threeparttable}
\begin{tabular}{cccc}
\hline
\hline
\textbf{Mass flow rate}& \textbf{Kn = 0.00181} & \textbf{Kn = 0.0181}& \textbf{Kn = 0.181} \\
\hline
DSMC    & 1.40&   1.29&   0.866\\
SG13-MGKS    & 1.40&   1.27&   0.880   \\
Relative Error    & 0.0\%&   1.55\%&   1.62\%   \\
\hline
\hline
\end{tabular}
\label{tb3}
\end{threeparttable}       
\end{table}

\begin{table}[H]
\centering
\setlength{\abovecaptionskip}{10pt}
\setlength{\belowcaptionskip }{10pt}
\caption{Computational times and memory consumption of the pressure-driven flow through a pipe by various methods.}
\setlength{\tabcolsep}{5.5mm}
\begin{threeparttable}
\begin{tabular}{ccccc}
\hline
\hline
\textbf{}& \textbf{IDVM}& \textbf{G13-MGKS}& \textbf{SG13-MGKS}& \textbf{Speed Ratio}\\
\hline
\textbf{Time}& 6.35 hours&   0.56 hours& 0.13 hours&    48.9\\
\textbf{Memory}& 2560 MB&   169 MB&  53 MB&  10.1\\
\hline
\hline
\end{tabular}
\label{tb4}
\end{threeparttable}
\end{table}

\section{CONCLUSIONS}
\label{S:4}

In this research, we introduce a simplified version of the Grad's 13 moments distribution function-based moment gas kinetic solver (SG13-MGKS), tailored for efficient computation across a different flow regimes, spanning from near-continuum to moderately rarefied conditions. To ensure the correct Prandtl numbers, we incorporate the Shakhov collision model and validate its accuracy through simulations of Couette flow with temperature differences. Furthermore, with the aim of facilitating SG13-MGKS application on unstructured meshes, we explore strategies to simplify numerical flux computations, thereby minimizing the complexity associated with gradient calculations that involve intricate integral coefficients. We assess the performance of SG13-MGKS on both triangular and quadrilateral meshes for flow scenarios involving the NACA0012 airfoil and pressure-driven pipe flow, respectively. Our findings reveal that SG13-MGKS can achieve reasonably computational results for Knudsen numbers below 0.2. Notably, it demonstrates a remarkable enhancement in computational efficiency, outperforming conventional discrete velocity method which necessitate velocity space discretization. SG13-MGKS achieves computational efficiency improvements exceeding a hundredfold and substantially reduces memory consumption, reveal its potential for practical applications in various flow regimes on the unstructured meshes.

\section*{ACKNOWLEDGMENTS}
\label{S:5}

This work was partly supported by the Ministry of Education (MOE) of Singapore (Grant No. MOE2018-T2-1-135). We acknowledge the computing support provided by the National Super-computing Centre (NSCC) Singapore for Computational Science and the High-Performance Computing of NUS (NUS-HPC).

\appendix
\section{COMPUTATION FOR INTEGRATION PARAMETERS}
\label{A:1}

Here the integration conducted at the left side of the cell interface is taken as an example and the calculation of the integration on the right side only needs to replace the $\langle\cdot\rangle_{>0}$ by the $\langle\cdot\rangle_{<0}$. The parameters of $\left\langle\xi_{n}^{o} \xi_{\tau}^{p} \zeta^{q} \right\rangle_{>0}$ could be calculated by using binomial theory as follows

\begin{equation}
\begin{aligned}
\left\langle\xi_{n}^{o} \xi_{\tau}^{p} \zeta^{q} \right\rangle_{>0}=\sum_{m=0}^{p} \frac{p !}{m !(p-m) !}\left\langle\xi_{n}^{o} C_{\tau}^{m} \zeta^{q} \right\rangle_{>0}\left(U_{n}\right)^{p-m},
\end{aligned}
\label{eqA0}
\end{equation}

The moment of VDF $\left\langle\xi_{n}^{o} C_{\tau}^{p} \zeta^{q} \right\rangle_{>0}$ could computed as

\begin{equation}
\begin{aligned}
\left\langle\xi_{n}^{o} C_{\tau}^{p} \zeta^{q} \right\rangle_{>0}=&\left\langle\xi_{n}^{o} C_{n}^{0}\right\rangle_{>0}^{eq}\left\langle C_{\tau}^{p}\right\rangle^{eq}\left\langle\zeta^{q}\right\rangle^{eq}+\sigma_{n n}^{*}\left\langle\xi_{n}^{o} C_{n}^{2}\right\rangle_{>0}^{eq}\left\langle C_{\tau}^{p}\right\rangle^{eq}\left\langle\zeta^{q}\right\rangle^{eq} \\
&+2 \sigma_{n \tau}^{*}\left\langle\xi_{n}^{o} C_{n}^{1}\right\rangle_{>0}^{eq}\left\langle C_{\tau}^{p+1}\right\rangle^{eq}\left\langle\zeta^{q}\right\rangle^{eq}+\sigma_{\tau \tau}^{*}\left\langle\xi_{n}^{o} C_{n}^{0}\right\rangle_{>0}^{eq}\left\langle C_{\tau}^{p+2}\right\rangle^{eq}\left\langle\zeta^{q}\right\rangle^{eq} \\
&-\left(\sigma_{n n}^{*}+\sigma_{\tau \tau}^{*}\right)\left\langle\xi_{n}^{o} C_{n}^{0}\right\rangle_{>0}^{eq}\left\langle C_{\tau}^{p}\right\rangle^{eq}\left\langle\zeta^{q+2}\right\rangle^{eq} \\
&-q_{n}^{*}\left\langle\xi_{n}^{o} C_{n}^{1}\right\rangle_{>0}^{eq}\left\langle C_{\tau}^{p}\right\rangle^{eq}\left\langle\zeta^{q}\right\rangle^{eq}-q_{\tau}^{*}\left\langle\xi_{n}^{o} C_{n}^{0}\right\rangle_{>0}^{eq}\left\langle C_{\tau}^{p+1}\right\rangle^{eq}\left\langle\zeta^{q}\right\rangle^{eq} \\
&+0.4 \vartheta q_{n}^{*}\left(\left\langle\xi_{n}^{o} C_{n}^{3}\right\rangle_{>0}^{eq}\left\langle C_{\tau}^{p}\right\rangle^{eq}\left\langle\zeta^{q}\right\rangle^{eq}+\left\langle\xi_{n}^{o} C_{n}^{1}\right\rangle_{>0}^{eq}\left\langle C_{\tau}^{p+2}\right\rangle^{eq}\left\langle\zeta^{q}\right\rangle^{eq}\right.\\
&\left.+\left\langle\xi_{n}^{o} C_{n}^{1}\right\rangle_{>0}^{eq}\left\langle C_{\tau}^{p}\right\rangle^{eq}\left\langle\zeta^{q+2}\right\rangle^{eq}\right)+0.4 \vartheta q_{\tau}^{*}\left(\left\langle\xi_{n}^{o} C_{n}^{2}\right\rangle_{>0}^{eq}\left\langle C_{\tau}^{p+1}\right\rangle^{eq}\left\langle\zeta^{q}\right\rangle^{eq}\right.\\
&\left.+\left\langle\xi_{n}^{o} C_{n}^{0}\right\rangle_{>0}^{eq}\left\langle C_{\tau}^{p+3}\right\rangle^{eq}\left\langle\zeta^{q}\right\rangle^{eq}+\left\langle\xi_{n}^{o} C_{n}^{0}\right\rangle_{>0}^{eq}\left\langle C_{\tau}^{p+1}\right\rangle^{eq}\left\langle\zeta^{q+2}\right\rangle^{eq}\right),
\end{aligned}
\label{A1}
\end{equation}

\noindent where $\vartheta=1 /\left(2 R_{g} T\right)$. The terms of stress and heat flux marked with an asterisk superscript are given by

\begin{equation}
\sigma_{n n}^{*}=\sigma_{n n} /(2 p R T), \quad \sigma_{n \tau}^{*}=\sigma_{n \tau} /(2 p R T), \quad \sigma_{\tau \tau}^{*}=\sigma_{\tau \tau} /(2 p R T),
\label{A2}
\end{equation}

\noindent and

\begin{equation}
q_{n}^{*}=q_{n} /(p R T), \quad q_{\tau}^{*}=q_{\tau} /(p R T).
\label{A3}
\end{equation}

\noindent The next task is to calculate terms of $\left\langle\xi_{n}^{p} C_{n}^{q}\right\rangle_{>0}^{eq}$, $\left\langle C_{\tau}^{k}\right\rangle^{eq}$ and $\left\langle\zeta^{k}\right\rangle^{eq}$, which are related to the moment integration of equilibrium state. Also from the binomial theory, $\left\langle\xi_{n}^{p} C_{n}^{q}\right\rangle_{>0}^{eq}$ could be expressed by the linear combination of $\left\langle\xi_{n}^{k}\right\rangle_{>0}^{eq}$ as

\begin{equation}
\left\langle\xi_{n}^{p} C_{n}^{q}\right\rangle_{>0}^{eq}= \sum_{m=0}^{q} (-1)^{q-m} \frac{q !}{m !(q-m)!}\left\langle\xi_{n}^{m+p}\right\rangle_{>0}^{e q}\left(U_{n}^{L}\right)^{q-m}.
\label{A4}
\end{equation}

Considering that the expression of $\left\langle\xi_{n}^{k}\right\rangle_{>0}^{eq}$ and $\left\langle\xi_{n}^{k}\right\rangle_{<0}^{eq}$ have different manner, the integration parameters related to the equilibrium state $\left\langle C_{\tau}^{k}\right\rangle^{eq}$, $\left\langle\zeta^{k}\right\rangle^{eq}$, $\left\langle\xi_{n}^{k}\right\rangle_{>0}^{eq}$ and $\left\langle\xi_{n}^{k}\right\rangle_{<0}^{eq}$ are given in \ref{B:1}.

\section{COMPUTATION OF INTEGRATION PARAMETERS RELATED TO THE EQUILIBRIUM STATE}
\label{B:1}

Taking the integral from zero to infinite on the left side of the cell interface, the parameter $\left\langle\xi_{n}^{k}\right\rangle_{>0}^{eq}$ could be given as

\begin{equation}
\left\langle\xi_{n}^{0}\right\rangle_{>0}^{e q}=\frac{1}{2}\left[1+\operatorname{erf}\left(\sqrt{\vartheta^{L}} U_{n}^{L}\right)\right],
\label{eqB1}
\end{equation}

\begin{equation}
\left\langle\xi_{n}^{1}\right\rangle_{>0}^{e q}=U_{n}^{L}\left\langle\xi_{n}^{0}\right\rangle_{>0}^{e q}+\frac{1}{2} \frac{e^{-\vartheta^{L}\left(U_{n}^{L}\right)^{2}}}{\sqrt{\vartheta^{L} \pi}},
\label{eqB2}
\end{equation}

\begin{equation}
\left\langle\xi_{n}^{k+2}\right\rangle_{>0}^{e q}=U_{n}^{L}\left\langle\xi_{n}^{k+1}\right\rangle_{>0}^{e}+\frac{k+1}{2 \vartheta^{L}}\left\langle\xi_{n}^{k}\right\rangle_{>0}^{e q}, \quad k=0,1,2, \ldots,
\label{eqB3}
\end{equation}

\noindent Similarly, taking the integral from negative infinite to zero, the parameter $\left\langle\xi_{n}^{k}\right\rangle_{<0}^{eq}$ are 

\begin{equation}
\left\langle\xi_{n}^{0}\right\rangle_{<0}^{e q}=\frac{1}{2} \operatorname{erfc}\left(\sqrt{\vartheta^{R}} U_{n}^{R}\right),
\label{eqB4}
\end{equation}

\begin{equation}
\left\langle\xi_{n}^{1}\right\rangle_{<0}^{e q}=U_{n}^{R}\left\langle\xi_{n}^{0}\right\rangle_{<0}^{e q}-\frac{1}{2} \frac{e^{-\vartheta^{R}\left(U_{n}^{R}\right)^{2}}}{\sqrt{\vartheta^{R}\pi}},
\label{eqB5}
\end{equation}

\begin{equation}
\left\langle\xi_{n}^{k+2}\right\rangle_{<0}^{e q}=U_{n}^{R}\left\langle\xi_{n}^{k+1}\right\rangle_{<0}^{e q}+\frac{k+1}{2 \vartheta^{R}}\left\langle\xi_{n}^{k}\right\rangle_{<0}^{e q}, \quad k=0,1,2, \ldots,
\label{eqB6}
\end{equation}

Following the binomial theory, part of even order of integration parameters $\left\langle C_{\tau}^{k}\right\rangle^{eq}$ and $\left\langle\zeta^{k}\right\rangle^{eq}$ could be computed as

\begin{equation}
\left\langle C_{n}^{0}\right\rangle^{e q}=\left\langle C_{\tau}^{0}\right\rangle^{e q}=\left\langle\zeta^{0}\right\rangle^{e q}=1,
\label{eqB7}
\end{equation}

\begin{equation}
\left\langle C_{n}^{2}\right\rangle^{e q}=\left\langle C_{\tau}^{2}\right\rangle^{e q}=\left\langle\zeta^{2}\right\rangle^{e q}=\frac{1}{2 \vartheta},
\label{eqB8}
\end{equation}

\begin{equation}
\left\langle C_{n}^{4}\right\rangle^{e q}=\left\langle C_{\tau}^{4}\right\rangle^{e q}=\left\langle\zeta^{4}\right\rangle^{e q}=\frac{3}{4 \vartheta^{2}},
\label{eqB9}
\end{equation}

\begin{equation}
\left\langle C_{n}^{6}\right\rangle^{e q}=\left\langle C_{\tau}^{6}\right\rangle^{e q}=\left\langle\zeta^{6}\right\rangle^{e q}=\frac{15}{8 \vartheta^{3}},
\label{eqB10}
\end{equation}

\begin{equation}
\left\langle C_{n}^{6}\right\rangle^{e q}=\left\langle C_{\tau}^{6}\right\rangle^{e q}=\left\langle\zeta^{8}\right\rangle^{e q}=\frac{105}{16 \vartheta^{4}}.
\label{eqB11}
\end{equation}

\noindent When the order $k$ is odd, the moment integrals of $\left\langle C^{k}\right\rangle^{eq}$ and $\left\langle\zeta^{k}\right\rangle^{eq}$ are all zero.

\section{EXPRESSION OF INTEGRATION PARAMETERS  RELATED TO THE STRESS AND HEAT FLUX}
\label{C:1}

The formulations of parameters including $\mathbf{Y}(1) \sim \mathbf{Y}(4)$ can be computed as

\begin{equation}
\begin{aligned}
& \mathbf{Y}^L(1)=\left\langle\xi_n^2 \xi_\tau^0 \zeta^0\right\rangle_{>0}, \mathbf{Y}^R(1)=\left\langle\xi_n^2 \xi_\tau^0 \zeta^0\right\rangle_{<0}, \\
& \mathbf{Y}^L(2)=\left\langle\xi_n^1 \xi_\tau^1 \zeta^0\right\rangle_{>0}, \mathbf{Y}^R(2)=\left\langle\xi_n^1 \xi_\tau^1 \zeta^0\right\rangle_{<0}, \\
& \mathbf{Y}^L(3)=\left\langle\xi_n^0 \xi_\tau^2 \zeta^0\right\rangle_{>0}, \mathbf{Y}^R(3)=\left\langle\xi_n^0 \xi_\tau^2 \zeta^0\right\rangle_{<0}, \\
& \mathbf{Y}^L(3)=\left\langle\xi_n^0 \xi_\tau^2 \zeta^0\right\rangle_{>0}, \mathbf{Y}^R(3)=\left\langle\xi_n^0 \xi_\tau^2 \zeta^0\right\rangle_{<0} .
\end{aligned}
\end{equation}

Formulations of parameters including $\mathbf{Z}(1) \sim \mathbf{Z}(6)$ can be computed as

\begin{equation}
\begin{aligned}
& \mathbf{Z}^L(1)=\left\langle\xi_n^4 \xi_\tau^0 \zeta^0\right\rangle_{>0}, \mathbf{Z}^R(1)=\left\langle\xi_n^4 \xi_\tau^0 \zeta^0\right\rangle_{<0}, \\
& \mathbf{Z}^L(2)=\left\langle\xi_n^2 \xi_\tau^2 \zeta^0\right\rangle_{>0}, \mathbf{Z}^R(2)=\left\langle\xi_n^2 \xi_\tau^2 \zeta^0\right\rangle_{<0}, \\
& \mathbf{Z}^L(3)=\left\langle\xi_n^2 \xi_\tau^0 \zeta^2\right\rangle_{>0}, \mathbf{Z}^R(3)=\left\langle\xi_n^2 \xi_\tau^0 \zeta^2\right\rangle_{<0}, \\
& \mathbf{Z}^L(4)=\left\langle\xi_n^3 \xi_\tau^1 \zeta^0\right\rangle_{>0}, \mathbf{Z}^R(4)=\left\langle\xi_n^3 \xi_\tau^1 \zeta^0\right\rangle_{<0}, \\
& \mathbf{Z}^L(5)=\left\langle\xi_n^1 \xi_\tau^3 \zeta^0\right\rangle_{>0}, \mathbf{Z}^R(5)=\left\langle\xi_n^1 \xi_\tau^3 \zeta^0\right\rangle_{<0} \\
& \mathbf{Z}^L(6)=\left\langle\xi_n^1 \xi_\tau^1 \zeta^2\right\rangle_{>0}, \mathbf{Z}^R(6)=\left\langle\xi_n^1 \xi_\tau^1 \zeta^2\right\rangle_{<0} .
\end{aligned}
\end{equation}


\bibliographystyle{reference_style.bst}

\begin{thebibliography}{33}
\expandafter\ifx\csname natexlab\endcsname\relax\def\natexlab#1{#1}\fi
\providecommand{\bibinfo}[2]{#2}
\ifx\xfnm\relax \def\xfnm[#1]{\unskip,\space#1}\fi
\bibitem[{Li and Zhang(2009)}]{li_gas-kinetic_2009}
\bibinfo{author}{Z.~H. Li}, \bibinfo{author}{H.~X. Zhang},
\newblock \bibinfo{title}{Gas-kinetic numerical studies of three-dimensional
  complex flows on spacecraft re-entry},
\newblock \bibinfo{journal}{Journal of Computational Physics}
  \bibinfo{volume}{228} (\bibinfo{year}{2009}) \bibinfo{pages}{1116--1138}.
\bibitem[{Li et~al.(2011)Li, Bi, Zhang, and Li}]{li_gas-kinetic_2011}
\bibinfo{author}{Z.-H. Li}, \bibinfo{author}{L.~Bi}, \bibinfo{author}{H.-X.
  Zhang}, \bibinfo{author}{L.~Li},
\newblock \bibinfo{title}{Gas-kinetic numerical study of complex flow problems
  covering various flow regimes},
\newblock \bibinfo{journal}{Computers \& Mathematics with Applications}
  \bibinfo{volume}{61} (\bibinfo{year}{2011}) \bibinfo{pages}{3653--3667}.
\bibitem[{Chen et~al.(2020)Chen, Zhu, and Xu}]{chen_three-dimensional_2020}
\bibinfo{author}{Y.~Chen}, \bibinfo{author}{Y.~Zhu}, \bibinfo{author}{K.~Xu},
\newblock \bibinfo{title}{A three-dimensional unified gas-kinetic wave-particle
  solver for flow computation in all regimes},
\newblock \bibinfo{journal}{Physics of Fluids} \bibinfo{volume}{32}
  (\bibinfo{year}{2020}) \bibinfo{pages}{096108}.
\bibitem[{Wang et~al.(2022)Wang, Liu, Zhuo, and
  Zhong}]{wang_investigation_2022}
\bibinfo{author}{Y.~Wang}, \bibinfo{author}{S.~Liu}, \bibinfo{author}{C.~Zhuo},
  \bibinfo{author}{C.~Zhong},
\newblock \bibinfo{title}{Investigation of nonlinear squeeze-film damping
  involving rarefied gas effect in micro-electro-mechanical systems},
\newblock \bibinfo{journal}{Computers \& Mathematics with Applications}
  \bibinfo{volume}{114} (\bibinfo{year}{2022}) \bibinfo{pages}{188--209}.
\bibitem[{Su et~al.(2017)Su, Liu, Zhang, and Wu}]{su_rarefaction_2017}
\bibinfo{author}{W.~Su}, \bibinfo{author}{H.~Liu}, \bibinfo{author}{Y.~Zhang},
  \bibinfo{author}{L.~Wu},
\newblock \bibinfo{title}{Rarefaction cloaking: {Influence} of the fractal
  rough surface in gas slider bearings},
\newblock \bibinfo{journal}{Physics of Fluids} \bibinfo{volume}{29}
  (\bibinfo{year}{2017}) \bibinfo{pages}{102003}.
\bibitem[{Zeng et~al.(2023)Zeng, Yuan, Zhang, Li, and Wu}]{zeng_general_2023-1}
\bibinfo{author}{J.~Zeng}, \bibinfo{author}{R.~Yuan},
  \bibinfo{author}{Y.~Zhang}, \bibinfo{author}{Q.~Li}, \bibinfo{author}{L.~Wu},
\newblock \bibinfo{title}{General synthetic iterative scheme for polyatomic
  rarefied gas flows},
\newblock \bibinfo{journal}{Computers \& Fluids} \bibinfo{volume}{265}
  (\bibinfo{year}{2023}) \bibinfo{pages}{105998}.
\bibitem[{Tantos et~al.(2020)Tantos, Varoutis, and
  Day}]{tantos_deterministic_2020}
\bibinfo{author}{C.~Tantos}, \bibinfo{author}{S.~Varoutis},
  \bibinfo{author}{C.~Day},
\newblock \bibinfo{title}{Deterministic and stochastic modeling of rarefied gas
  flows in fusion particle exhaust systems},
\newblock \bibinfo{journal}{Journal of Vacuum Science \& Technology B,
  Nanotechnology and Microelectronics: Materials, Processing, Measurement, and
  Phenomena} \bibinfo{volume}{38} (\bibinfo{year}{2020})
  \bibinfo{pages}{064201}.
\bibitem[{Liu et~al.(2023{\natexlab{a}})Liu, Yang, Zhang, Teo, and
  Shu}]{liu_simplified_2023}
\bibinfo{author}{W.~Liu}, \bibinfo{author}{L.~Yang},
  \bibinfo{author}{Z.~Zhang}, \bibinfo{author}{C.~Teo},
  \bibinfo{author}{C.~Shu},
\newblock \bibinfo{title}{Simplified hydrodynamic-wave particle method for the
  multiscale rarefied flow},
\newblock \bibinfo{journal}{Applied Mathematical Modelling}
  \bibinfo{volume}{116} (\bibinfo{year}{2023}{\natexlab{a}})
  \bibinfo{pages}{469--489}.
\bibitem[{Liu et~al.(2023{\natexlab{b}})Liu, Shi, and Liu}]{liu_numerical_2023}
\bibinfo{author}{W.~Liu}, \bibinfo{author}{L.~Shi}, \bibinfo{author}{H.~Liu},
\newblock \bibinfo{title}{Numerical study of the impact of geometrical
  parameters on the rarefied gas transport in porous media},
\newblock \bibinfo{journal}{Gas Science and Engineering} \bibinfo{volume}{110}
  (\bibinfo{year}{2023}{\natexlab{b}}) \bibinfo{pages}{204855}.
\bibitem[{Wu et~al.(2016)Wu, Liu, Reese, and Zhang}]{wu_non-equilibrium_2016}
\bibinfo{author}{L.~Wu}, \bibinfo{author}{H.~Liu}, \bibinfo{author}{J.~M.
  Reese}, \bibinfo{author}{Y.~Zhang},
\newblock \bibinfo{title}{Non-equilibrium dynamics of dense gas under tight
  confinement},
\newblock \bibinfo{journal}{Journal of Fluid Mechanics} \bibinfo{volume}{794}
  (\bibinfo{year}{2016}) \bibinfo{pages}{252--266}. \bibinfo{note}{Publisher:
  Cambridge University Press}.
\bibitem[{Xu(2021)}]{xu_unified_2021-1}
\bibinfo{author}{K.~Xu}, \bibinfo{title}{A {Unified} {Computational} {Fluid}
  {Dynamics} {Framework} from {Rarefied} to {Continuum} {Regimes}},
  \bibinfo{publisher}{Cambridge University Press}, \bibinfo{edition}{1}
  edition, \bibinfo{year}{2021}.
\bibitem[{Zhu et~al.(2019)Zhu, Liu, Zhong, and Xu}]{zhu_unified_2019}
\bibinfo{author}{Y.~Zhu}, \bibinfo{author}{C.~Liu}, \bibinfo{author}{C.~Zhong},
  \bibinfo{author}{K.~Xu},
\newblock \bibinfo{title}{Unified gas-kinetic wave-particle methods. {II}.
  {Multiscale} simulation on unstructured mesh},
\newblock \bibinfo{journal}{Physics of Fluids} \bibinfo{volume}{31}
  (\bibinfo{year}{2019}) \bibinfo{pages}{067105}.
\bibitem[{Bird(1994)}]{bird_molecular_1994}
\bibinfo{author}{G.~Bird}, \bibinfo{title}{Molecular {Gas} {Dynamics} and the
  {Direct} {Simulation} of {Gas} {Flows}}, \bibinfo{publisher}{Oxford
  University Press}, \bibinfo{year}{1994}.
\bibitem[{Goldstein et~al.(1989)Goldstein, Sturtevant, and
  Broadwell}]{goldstein_investigations_1989}
\bibinfo{author}{D.~Goldstein}, \bibinfo{author}{B.~Sturtevant},
  \bibinfo{author}{J.~E. Broadwell},
\newblock \bibinfo{title}{Investigations of the motion of discrete-velocity
  gases},
\newblock in: \bibinfo{booktitle}{Progress in {Astronautics} and
  {Aeronautics}}, \bibinfo{publisher}{vol. 118, AIAA},
  \bibinfo{address}{Washington}, \bibinfo{year}{1989}, p. \bibinfo{pages}{202}.
\bibitem[{Harris(2021)}]{harris_solution-adaptive_2021}
\bibinfo{author}{R.~E. Harris},
\newblock \bibinfo{title}{Solution-{Adaptive} {Boltzmann} {Discrete} {Velocity}
  {Method} for {Rarefied} {Flows} in {Diatomic} {Gases}},
\newblock in: \bibinfo{booktitle}{{AIAA} {SCITECH} 2022 {Forum}}, {AIAA}
  {SciTech} {Forum}, \bibinfo{publisher}{American Institute of Aeronautics and
  Astronautics}, \bibinfo{year}{2021}, p.~\bibinfo{pages}{1}.
\bibitem[{Grad(1949)}]{grad_kinetic_1949}
\bibinfo{author}{H.~Grad},
\newblock \bibinfo{title}{On the kinetic theory of rarefied gases},
\newblock \bibinfo{journal}{Communications on Pure and Applied Mathematics}
  \bibinfo{volume}{2} (\bibinfo{year}{1949}) \bibinfo{pages}{331--407}.
\bibitem[{Grad(1952)}]{grad_statistical_1952}
\bibinfo{author}{H.~Grad},
\newblock \bibinfo{title}{Statistical mechanics, thermodynamics, and fluid
  dynamics of systems with an arbitrary number of integrals},
\newblock \bibinfo{journal}{Communications on Pure and Applied Mathematics}
  \bibinfo{volume}{5} (\bibinfo{year}{1952}) \bibinfo{pages}{455--494}.
\bibitem[{Struchtrup(2005)}]{struchtrup_macroscopic_2005}
\bibinfo{author}{H.~Struchtrup}, \bibinfo{title}{Macroscopic transport
  equations for rarefied gas flows: approximation methods in kinetic theory},
  Interaction of mechanics and mathematics series,
  \bibinfo{publisher}{Springer}, \bibinfo{address}{Berlin ; New York},
  \bibinfo{year}{2005}. \bibinfo{note}{OCLC: ocm60846281}.
\bibitem[{Gu and Emerson(2009)}]{gu_high-order_2009}
\bibinfo{author}{X.~J. Gu}, \bibinfo{author}{D.~R. Emerson},
\newblock \bibinfo{title}{A high-order moment approach for capturing
  non-equilibrium phenomena in the transition regime},
\newblock \bibinfo{journal}{Journal of Fluid Mechanics} \bibinfo{volume}{636}
  (\bibinfo{year}{2009}) \bibinfo{pages}{177--216}.
\bibitem[{Wu and Gu(2020)}]{wu_accuracy_2020}
\bibinfo{author}{L.~Wu}, \bibinfo{author}{X.~J. Gu},
\newblock \bibinfo{title}{On the accuracy of macroscopic equations for
  linearized rarefied gas flows},
\newblock \bibinfo{journal}{Advances in Aerodynamics} \bibinfo{volume}{2}
  (\bibinfo{year}{2020}) \bibinfo{pages}{2}.
\bibitem[{Liu et~al.(2023)Liu, Liu, Zhang, Teo, and Shu}]{liu_grads_2023}
\bibinfo{author}{W.~Liu}, \bibinfo{author}{Z.~J. Liu}, \bibinfo{author}{Z.~L.
  Zhang}, \bibinfo{author}{C.~J. Teo}, \bibinfo{author}{C.~Shu},
\newblock \bibinfo{title}{Grad's distribution function for 13 moments-based
  moment gas kinetic solver for steady and unsteady rarefied flows: {Discrete}
  and explicit forms},
\newblock \bibinfo{journal}{Computers \& Mathematics with Applications}
  \bibinfo{volume}{137} (\bibinfo{year}{2023}) \bibinfo{pages}{112--125}.
\bibitem[{Chapman and Cowling(1962)}]{chapman_mathematical_1962}
\bibinfo{author}{S.~Chapman}, \bibinfo{author}{T.~G. Cowling},
\newblock \bibinfo{title}{The {Mathematical} {Theory} of {Non}-{Uniform}
  {Gases}},
\newblock \bibinfo{journal}{American Journal of Physics} \bibinfo{volume}{30}
  (\bibinfo{year}{1962}) \bibinfo{pages}{389--389}.
\bibitem[{Pekardan and Alexeenko(2018)}]{pekardan_rarefaction_2018}
\bibinfo{author}{C.~Pekardan}, \bibinfo{author}{A.~Alexeenko},
\newblock \bibinfo{title}{Rarefaction {Effects} for {Transonic} {Airfoil}
  {Flows} at {Low} {Reynolds} {Numbers}},
\newblock \bibinfo{journal}{AIAA Journal} \bibinfo{volume}{56}
  (\bibinfo{year}{2018}) \bibinfo{pages}{765--779}. \bibinfo{note}{Publisher:
  American Institute of Aeronautics and Astronautics}.
\bibitem[{Shakhov(1968)}]{shakhov_generalization_1968}
\bibinfo{author}{E.~M. Shakhov},
\newblock \bibinfo{title}{Generalization of the {Krook} kinetic relaxation
  equation},
\newblock \bibinfo{journal}{Fluid Dynamics} \bibinfo{volume}{3}
  (\bibinfo{year}{1968}) \bibinfo{pages}{95--96}.
\bibitem[{Xu and Huang(2010)}]{xu_unified_2010}
\bibinfo{author}{K.~Xu}, \bibinfo{author}{J.-C. Huang},
\newblock \bibinfo{title}{A unified gas-kinetic scheme for continuum and
  rarefied flows},
\newblock \bibinfo{journal}{Journal of Computational Physics}
  \bibinfo{volume}{229} (\bibinfo{year}{2010}) \bibinfo{pages}{7747--7764}.
\bibitem[{Yang et~al.(2018)Yang, Shu, Yang, Chen, and
  Dong}]{yang_improved_2018}
\bibinfo{author}{L.~M. Yang}, \bibinfo{author}{C.~Shu}, \bibinfo{author}{W.~M.
  Yang}, \bibinfo{author}{Z.~Chen}, \bibinfo{author}{H.~Dong},
\newblock \bibinfo{title}{An improved discrete velocity method ({DVM}) for
  efficient simulation of flows in all flow regimes},
\newblock \bibinfo{journal}{Physics of Fluids} \bibinfo{volume}{30}
  (\bibinfo{year}{2018}) \bibinfo{pages}{062005}.
\bibitem[{Liu et~al.(2024)Liu, Zhang, Zeng, and Wu}]{liu_further_2024}
\bibinfo{author}{W.~Liu}, \bibinfo{author}{Y.~Zhang},
  \bibinfo{author}{J.~Zeng}, \bibinfo{author}{L.~Wu},
\newblock \bibinfo{title}{Further acceleration of multiscale simulation of
  rarefied gas flow via a generalized boundary treatment},
\newblock \bibinfo{journal}{Journal of Computational Physics}
  \bibinfo{volume}{503} (\bibinfo{year}{2024}) \bibinfo{pages}{112830}.
\bibitem[{Grünfeld and Marinescu(2014)}]{grunfeld_time_2014}
\bibinfo{author}{C.~Grünfeld}, \bibinfo{author}{D.~Marinescu},
\newblock \bibinfo{title}{On a time and space discretized approximation of the
  {Boltzmann} equation in the whole space},
\newblock \bibinfo{journal}{Computers \& Mathematics with Applications}
  \bibinfo{volume}{68} (\bibinfo{year}{2014}) \bibinfo{pages}{1393--1408}.
\bibitem[{Venkatakrishnan(1995)}]{venkatakrishnan_convergence_1995}
\bibinfo{author}{V.~Venkatakrishnan},
\newblock \bibinfo{title}{Convergence to {Steady} {State} {Solutions} of the
  {Euler} {Equations} on {Unstructured} {Grids} with {Limiters}},
\newblock \bibinfo{journal}{Journal of Computational Physics}
  \bibinfo{volume}{118} (\bibinfo{year}{1995}) \bibinfo{pages}{120--130}.
\bibitem[{Bjorck(1990)}]{bjorck_least_1990}
\bibinfo{author}{A.~Bjorck},
\newblock \bibinfo{title}{Least squares methods},
\newblock in: \bibinfo{booktitle}{Handbook of {Numerical} {Analysis}},
  volume~\bibinfo{volume}{1}, \bibinfo{publisher}{Elsevier},
  \bibinfo{year}{1990}, pp. \bibinfo{pages}{465--652}.
\bibitem[{{L.M. Yang} et~al.(2019){L.M. Yang}, Shu, Yang, and
  Wu}]{lm_yang_improved_2019}
\bibinfo{author}{{L.M. Yang}}, \bibinfo{author}{C.~Shu},
  \bibinfo{author}{W.~Yang}, \bibinfo{author}{J.~Wu},
\newblock \bibinfo{title}{An improved three-dimensional implicit discrete
  velocity method on unstructured meshes for all {Knudsen} number flows},
\newblock \bibinfo{journal}{Journal of Computational Physics}
  \bibinfo{volume}{396} (\bibinfo{year}{2019}) \bibinfo{pages}{738--760}.
\bibitem[{Cercignani(1969)}]{cercignani_mathematical_1969}
\bibinfo{author}{C.~Cercignani}, \bibinfo{title}{Mathematical {Methods} in
  {Kinetic} {Theory}}, \bibinfo{publisher}{Springer US},
  \bibinfo{address}{Boston, MA}, \bibinfo{year}{1969}.
\bibitem[{Yang et~al.(2018)Yang, Chen, Shu, Yang, Wu, and
  Zhang}]{yang_implicit_2018}
\bibinfo{author}{L.~M. Yang}, \bibinfo{author}{Z.~Chen},
  \bibinfo{author}{C.~Shu}, \bibinfo{author}{W.~M. Yang},
  \bibinfo{author}{J.~Wu}, \bibinfo{author}{L.~Q. Zhang},
\newblock \bibinfo{title}{implicit {Improved} fully implicit discrete-velocity
  method for efficient simulation of flows in all flow regimes},
\newblock \bibinfo{journal}{Physical Review E} \bibinfo{volume}{98}
  (\bibinfo{year}{2018}) \bibinfo{pages}{063313}.

\end{thebibliography}

\end{document}